\documentclass[a4paper, 11pt]{article} 

\usepackage{aasmacros}

\usepackage[english]{babel}               

\usepackage{lmodern}        
\usepackage[T1]{fontenc}    
\usepackage{mathpazo}                                            
\usepackage[utf8]{inputenc} 

\usepackage[dvipsnames]{xcolor}

\usepackage{amsmath}      
\usepackage{amsfonts}     
\usepackage{amssymb}      
\usepackage{mathrsfs}
\usepackage[cal=boondox,scr=boondoxo]{mathalfa}

\usepackage[normalem]{ulem} %

\usepackage[italicdiff]{physics}
\usepackage{tabu}

\usepackage{geometry}     
\usepackage[final, babel]{microtype} 

\usepackage[textwidth=2.4cm]{todonotes}%

\usepackage{graphicx} 
\usepackage{float}%

\usepackage{csquotes} %

\usepackage{lipsum}

\usepackage{jcappub}

\DeclareFontFamily{U}{mathx}{\hyphenchar\font45}
\DeclareFontShape{U}{mathx}{m}{n}{
      <5> <6> <7> <8> <9> <10>
      <10.95> <12> <14.4> <17.28> <20.74> <24.88>
      mathx10
      }{}
\DeclareSymbolFont{mathx}{U}{mathx}{m}{n}
\DeclareMathSymbol{\bigtimes}{1}{mathx}{"91}

\newcommand{\code}[1]{\texttt{#1}} 
\newcommand{\obs}{\mathrm{obs}}
\newcommand{\Mobs}{M^{\obs}}

\newcommand{\sigmaobs}{\sigma_{\mu_\obs|\mu}}
\newcommand{\rhotot}{\rho_{\mathrm{tot}}}

\newcommand{\rhobary}{\rho_b}

\newcommand{\rhocdm}{\rho_\mathrm{CDM}}

\newcommand{\SigmaAvg}[1]{\ev{\Sigma(<#1)}}
\newcommand{\losDist}{x}

\newcommand{\twohaloA}{a_{2\mathrm{h}}}
\newcommand{\rhoref}{\rho_{\mathrm{ref}}}

\newcommand{\fullRadiiBaryResult}{-0.00705_{0.036}^{0.045}}
\newcommand{\fullRadiiBaryOneHaloOnlyResult}{-0.0231_{0.035}^{0.049}}
\newcommand{\fullRadiiNfwResult}{-0.0781_{0.020}^{0.021}}
\newcommand{\fullRadiiNfwOneHaloOnlyResult}{-0.0829_{0.019}^{0.021}}

\newcommand{\cutRadiiBaryResult}{-0.0102_{0.041}^{0.055}}
\newcommand{\cutRadiiBaryOneHaloOnlyResult}{-0.0311_{0.038}^{0.055}}
\newcommand{\cutRadiiNfwResult}{-0.0935_{0.021}^{0.023}}
\newcommand{\cutRadiiNfwOneHaloOnlyResult}{-0.0995_{0.021}^{0.021}}

\newcommand{\fullRadiiPctBiasBary}{-0.7\%}

\newcommand{\fullRadiiPctBiasNfw}{-7.5\%}

\newcommand{\cutRadiiPctBiasBaryOneHalo}{-3.1\%}

\newcommand{\cutRadiiPctBiasNfwOneHalo}{-9.5\%}

\newcommand{\fullRadiiBiasNumSigmaAwayBary}{0.17}

\newcommand{\fullRadiiBiasNumSigmaAwayNfw}{3.8}

\newcommand{\cutRadiiBiasNumSigmaAwayBaryOneHalo}{0.63}

\newcommand{\cutRadiiBiasNumSigmaAwayNfwOneHalo}{4.8}

\newcommand{\abstractBaryBias}{0.7\%}
\newcommand{\abstractNfwBias}{7.5\%}

\graphicspath{ {./figs/} }

\title{Towards 1\% accurate galaxy cluster masses: Including baryons in weak-lensing mass inference}
\author[a]{Dylan Cromer,}
\author[a]{Nicholas Battaglia,}
\author[b, c]{Hironao Miyatake,}
\author[d]{and Melanie Simet}

\affiliation[a]{Department of Astronomy, Cornell University, Ithaca, NY 14853, USA}
\affiliation[b]{Kobayashi-Maskawa Institute for the Origin of Particles and the Universe (KMI),
Nagoya University, Nagoya, 464-8602, Japan}
\affiliation[c]{Kavli Institute for the Physics and Mathematics of the Universe (Kavli IPMU, WPI), UTIAS, The University of Tokyo, 5-1-5 Kashiwa-no-ha, Kashiwa, Chiba 277- 8583, Japan}
\affiliation[d]{Industry}

\emailAdd{dmc396@cornell.edu}

\date{\today}
\abstract{Galaxy clusters are a promising probe of late-time structure growth, but constraints on cosmology from cluster abundances are currently limited by systematics in their inferred masses. One unmitigated systematic effect in weak-lensing mass inference is ignoring the presence of baryons and treating the entire cluster as a dark matter halo. In this work we present a new flexible model for cluster densities that captures both the baryonic and dark matter profiles, a new general technique for calculating the lensing signal of an arbitrary density profile, and a methodology for stacking those lensing signal to appropriately model stacked weak-lensing measurements of galaxy cluster catalogues. We test this model on 1400 simulated clusters. Similarly to previous studies, we find that a dark matter-only model overestimates the average mass by $\abstractNfwBias{}$, but including our baryonic term reduces that to $\abstractBaryBias{}$. Additionally, to mitigate the computational complexity of our model, we construct an emulator (surrogate model) which accurately interpolates our model for parameter inference, while being much faster to use than the raw model. We also provide an open-source software framework for our model and emulator, called \code{maszcal}, which will serve as a platform for continued efforts to improve these mass-calibration techniques. In this work, we detail our model, the construction of the emulator, and the tests which we used to validate that our model does mitigate bias. Lastly, we describe tests of the emulator's accuracy.}

\begin{document}

\maketitle

\tableofcontents

\section{Introduction}

Galaxy clusters provide an excellent probe of the nonlinear regime of cosmology. Their abundance as a function of redshift is sensitive to the late-time growth of structure \citep{Allen2011, Weinberg2013}, which is in turn a promising frontier for exploring extensions to the six-parameter $\Lambda$CDM model of cosmology, such as dynamical dark energy ($w(z)$CDM) and a non-minimal neutrino mass sum ($\Sigma m_\nu$).

Galaxy clusters are identified across many wavelengths, commonly ranging from millimeters (microwave frequencies) to sub-nanometers (X-ray energies). Measurements of secondary temperature fluctuations in the CMB that arise from the thermal Sunyaev-Zel'dovich effect \citep[tSZ;][]{1970Ap&SS...7....3S} are emerging as a powerful tool to count clusters, since tSZ-selected cluster samples have well-behaved selection functions. Forecasts for Simons Observatory (SO) and CMB-S4 have already shown that tSZ cluster abundances provide independent and competitive constraints on dark energy and $\Sigma m_{\nu}$ \citep[e.g.,][]{LA2017, MBM2017, Cromer2019} compared to CMB lensing, galaxy clustering, and tomographic weak-lensing measurements. Another powerful method of selecting clusters is through the X-ray emissions of their hot ionized gas. eROSITA, the primary instrument on the Spectr-RG space mission, is currently observing the X-ray sky, and is expected to find $\sim 10^5$ clusters \citep{erosita-clusters-forecast}, providing an additional cutting-edge measurement of cluster abundances. Furthermore, new high fidelity weak-lensing measurements of around $20,000$ galaxy clusters will be made by the Vera Rubin Observatory \citep{lsst-summary} in the Legacy Survey of Space and Time (LSST). These measurements provide an independent measurement of cluster mass that is critical to using cluster abundances for dark energy and $\Sigma m_\nu$, as we will discuss below. The caveats that accompany such forecasts are systematic uncertainties that we need to quantify and mitigate. It is these systematic uncertainties that will be the biggest hurdle for SO, CMB-S4, eROSITA, and the Vera Rubin Observatory when they use cluster abundances to probe $\Sigma m_{\nu}$ and dark energy.

Currently, the largest systematic uncertainty in cluster cosmology is an accurate calibration of an observable-to-mass relation. The recent cosmological constraints from cluster abundances \citep[e.g.,][]{Vik2009, Hass2013, PlnkSZCos2015, Mantz2015, deHaan2016, Bocquet2018} are all limited in this way. For example, while SO and CMB-S4 will find on the order of $10^4$ to $10^5$ galaxy clusters through their tSZ signal \citep[e.g,][]{CMBS4, LA2017, MBM2017, Mantz2019}, the additional statistical power of these measurements will be wasted if the mass calibration accuracy is not improved.

The preferential observational technique used to calibrate cluster mass is weak-lensing, as it provides an independent method for measuring and calibrating cluster masses \citep{CMBS4, LSSTSRD}. The weak-lensing signal from galaxy clusters appears as small but coherent distortions (``shear'') in the background galaxy shapes that result from the gravitational deflection of light. This is the most direct probe of total cluster mass \citep[e.g.,][]{Bart2001, Ref2003} since it depends only on the gravitational potential sourced by both baryonic and dark matter, for a fixed lens redshift. Furthermore, N-body simulations have shown that weak-lensing mass measurements are unbiased with respect to the true mass \citep[e.g.,][]{Becker2011} in idealized situations. The technique of stacking weak-lensing signals of clusters, and fitting the parameters of the stacked profile, has the benefit of reducing the amount of triaxiality and sub-structure in the measurement \citep[e.g.,][]{Corless2009} and makes the errors more gaussian \citep[e.g.,][]{Simet2017}. However, weak-lensing mass calibration is not a panacea and there are differences that exist among published weak-lensing masses for identical clusters from independent analysis groups \citep[e.g.,][]{OS2016}. These differences are likely the results of underestimated systematics in the observations (e.g., blending of galaxies, photometric redshifts, or shape measurements). Current \emph{stage-3} imaging surveys like the Dark Energy Survey (DES), Hyper Suprime-Cam Survey (HSC), and Kilo-Degree Survey (KIDS) are tackling these observational systematic hurdles as they progress towards the level of precision necessary for Rubin Observatory measurement requirements, with theoretical efforts also being pursued towards this end \citep{grandis-etal-2021}. As the community makes progress on the observational systematics, a parallel effort is needed to mitigate systematic uncertainties associated with mass inferences (i.e., mass modeling), which will be critically important for the Rubin Observatory.

The fiducial model used to infer cluster masses from weak-lensing observations is the spherically symmetric NFW profile \citep{NFW1997}. This model accurately captures the shapes of the radial profile of dark matter halos across cosmologies, but it does not capture the shape of the baryon density close the center of galaxy clusters, due to baryon feedback. Because the amount of baryons in clusters is roughly at the cosmic mean  \citep[e.g.,][]{Gonzalez2013}, this can have significant impact on the overall cluster weak-lensing signal.

Initial models that attempted to account for baryonic effects on the total density profile would allow the concentration parameter to vary differently from a dark matter only prediction \citep[e.g.,][]{Rudd2008}. Physically such models were meant to account for the response of the dark matter halo to baryonic dissipation and would describe the mass profile to $\sim 10$ percent accuracy \citep[e.g.,][]{Gnedin2011}. This modeling approach was adopted in \cite{Shirasaki2018} for weak-lensing mass inferences and showed modest improvement in the cluster mass estimates, but do not reach the required sub-percent accuracy of the LSST.

Cosmological hydrodynamic simulations have clearly demonstrated that baryons bias the inferred weak-lensing mass \citep{Henson2017, Lee2018}. The bias found in simulations is mass dependent and a function of the sub-grid feedback model. What is more concerning is that the simple fix of allowing the concentration parameter to vary freely in order to capture these baryonic effects within the NFW model does not mitigate these biases \citep{Lee2018}. It is clear that a new model for the baryonic effects on the cluster weak-lensing signal is needed to reach the desired sub-percent mass calibration in order to place competitive constraints on dark energy and $\Sigma m_{\nu}$ from cluster abundances. A first attempt at such a model has been made using a fixed-shape baryon profile obtained from X-ray observations \citep{Debackere2021}.

In this paper, we provide the initial results from a general model that explicitly includes the baryonic density, represented by a Generalized-NFW (henceforth GNFW) profile \citep{Zhao96}. From this model of the density, we calculate the 1-halo excess surface density (henceforth ESD) $\Delta \Sigma$, which is proportional to the weak-lensing shear $\gamma$. When applying our model to radii large enough for the 2-halo density to be relevant, we include this density in our model. In Section \ref{sec:wl} we provide the details of how this model is calculated.

Additionally, we construct an emulator, or surrogate-model, for our modeled ESD. This emulator uses the techniques employed by the Coyote Universe Emulator \citep{coyote-universe-emulator} to create a highly accurate computational surrogate for use in Markov-Chain Monte Carlo (MCMC). This technique allows the full MCMC analysis routine of our tests to be run within an hour on a personal computer. In Section \ref{sec:emulator}, we describe the construction of this emulator, and in Appendix \ref{app:emulator-accuracy} we outline how we test this emulator for accuracy.

To test if and how well our model is able to mitigate the systematic bias from ignoring baryons, we use the \citep{battaglia-simulations} simulated clusters and weak-lensing ESD profiles. We calculate the average ESD profile for the simulation, then fit this using a stacked model, both with and without the baryonic density term. We examine the inferred bias parameter $a_\obs{}$, which is the natural logarithm of the average SZ-mass bias from the weak-lensing mass (see Equation \ref{eq:sz-mass-relation-noscatter} for an explicit definition), and in this test should be 0 when the model is an unbiased estimator of the mass.

In this paper, the simulation we test on uses $M_{500c}$ masses, so we adopt this mass definition for all of our results. However, the methodology and analysis software we have developed can be used with a general mass definition, using arbitrary $\Delta$ and either mean or critical reference density. At no point do we convert either mass or concentration to another mass definition such as $200m$. As a result, the concentrations we report are $c_{500c}$. Henceforth, unless otherwise noted, we write $500c$ masses and concentrations simply as $M$ and $c$ respectively.

\section{Weak Lensing Model} \label{sec:wl}

\subsection{Weak Lensing}

The most common estimator used to fit cluster weak-lensing data is the tangential shear $\gamma_t$ in the weak gravitational field limit. In this limit, the tangential shear is related to the line-of-sight projected density $\Sigma(r)$ by
\begin{align}
    \gamma_t(r) &= \frac{\ev{\Sigma(<r)} - \Sigma(r)}{\Sigma_{\mathrm{crit}}(z_{\mathrm{l}}, z_{\mathrm{s}})} \\
                &\equiv \frac{\Delta \Sigma(r)}{\Sigma_{\mathrm{crit}}(z_{\mathrm{l}}, z_{\mathrm{s}})},
\end{align}
where $\Delta\Sigma(r)$ is the excess surface density (ESD), and $\Sigma_{\mathrm{crit}}$ is the critical surface mass density given by
\begin{equation}
    \Sigma_{\mathrm{crit}}(z_{\mathrm{l}}, z_{\mathrm{s}}) = \frac{c^2}{4\pi G} \frac{D_A(z_s)}{(1+z_l)^2 D_A(z_l) D_A(z_l, z_s)},
\end{equation}
where $c$ is the speed of light, $G$ is the gravitational constant, $z_l$ is the lens (tSZ cluster) redshift, $z_s$ the source galaxy redshift, and $D_A$ is the angular diameter distance. The factor $(1+z_l)^{-2}$ is present due to our choice of $r$ as a comoving radius. We do not attempt to model the reduced shear, which is needed near the cluster center where the weak-field-limit begins to break down. This is not problematic for the purposes of this work; we compare our model directly to simulated ESD profiles which are not corrected for the effects of strong-lensing.

\subsection{Modeling Baryons}

Baryons account for $\sim 1/6\mathrm{th}$ the mass of galaxy clusters, and follow different radial distributions than pure cold dark matter halos. Not accounting for baryons in a weak-lensing mass estimate will bias the mass estimates \citep[e.g.,][]{Henson2017, Lee2018}. We address this by breaking the density into two terms:
\begin{equation}
    \rhotot{}(r) = \rhobary{}(r) + \rhocdm{}(r),
\end{equation}
where $\rhobary{}$ is a GNFW profile which models the baryons, and $\rhocdm{}$ is an NFW profile which models the dark matter, each weighted by the cosmic baryon fraction ($f_b = \Omega_b/\Omega_m$) and dark matter fraction ($1-f_b$) respectively:
\begin{align}
    \rhocdm(r) &= \frac{(1-f_b) \delta_c \rhoref{}}{x (1 + x)^2},
    \\
    \rhobary(r) &= \frac{f_b \rho_0}{y^\gamma \qty(1 + y^{1/\alpha})^{(\beta - \gamma)\alpha}},
\end{align}
where the GNFW parameters $\gamma$, $\alpha$, and $\beta$ are the cluster core power-law, transition power-law, and long-distance power-law respectively; $x$ is $r/r_s$, with $r_s$ the scale radius $r_\Delta/c$, $r_\Delta$ is the radius where the density is $\Delta$ times the mass definition reference density $\rhoref{}$, $y = x/x_c$ with $x_c = 0.5$ is a fixed core-scale, and $\delta_c$ is the usual characteristic halo overdensity as a function of concentration $c$ given by
\begin{equation}
    \delta_c = \frac{\Delta c^3}{3\qty[\ln(1+c) - c/(1+c)]}.
\end{equation}

For a general mass definition, $\Delta$ can have the subscripts $m$ for mean-matter density, or $c$ for critical density, the reference density $\rhoref{}$ will be either $\rho_m(z)/(1+z)^3$, or $\rho_c(z)/(1+z)^3$.\footnote{The factor $1/(1+z)^3$ arises from our use of comoving radii.} The coordinate $x$ is different between definitions as well, as it is defined to be
\begin{equation}
    x = c \qty(\frac{4 \pi \Delta \rhoref{}}{3 M})^{1/3} r,
\end{equation}
where $r$ is the comoving radius.

The factor $\rho_0$ ensures the baryon profile is normalized. GNFW profiles cannot be normalized analytically for arbitrary parameters, so instead we must numerically solve for $\rho_0$. We do this by enforcing that the baryon fraction must be preserved at a radius $R_b$:
\begin{equation}
    f_b = \frac{\int_{V(R_b)} \dd[3]{r} \rhobary(r)}{\int_{V(R_b)} \dd[3]{r} \rhotot(r)},
\end{equation}
where $V(R_b)$ is a sphere of radius $R_b$. Solving for $\rho_0$ results in
\begin{align}
    \rho_0 &= \frac{\int_0^{R_b} \dd{r} r^2 \qty(\delta_c \rho_\Delta) \qty(x (1 + x)^2)^{-1}}{\int_0^{R_b} \dd{r} r^2 y^{-\gamma} \qty(1 + y^{1/\alpha})^{-(\beta - \gamma)\alpha}}.
\end{align}
For the purposes of this work, we adopt a fixed $R_b$ of $3.3$ Mpc. This value is motivated by being an approximate upper-bound to where baryons should begin to trace the dark matter profile, but the exact value has no significance. To this end, we perform the same analysis with $R_b = 5$ Mpc, and find the results change insignificantly and our model is still unbiased.

There is no closed-form expression of $\Delta \Sigma$ for a GNFW profile in terms of standard functions, for arbitrary $\alpha, \beta, \gamma$. This necessitates that we numerically estimate $\Delta \Sigma$. To aid in this computation, we derive a new expression for $\Delta \Sigma$ valid for an arbitrary $\rho$, one which is merely the sum of two single integrations, eliminating the need to numerically estimate an integral over two variables. The details of this derivation are in Appendix \ref{app:rho_to_delta_sigma_calc}; the final expression is
\begin{equation}
    \Delta \Sigma(R) = \frac{4}{R^2} \int_0^R \dd{\losDist{}} \losDist{}^2 \rho(\losDist{}) - 4R \int_0^{\pi/2} \dd{\theta} \frac{\rho(R\sec{\theta})}{4\sin{\theta} + 3 - \cos(2\theta)}.
\end{equation}

\subsection{Modeling the 2-Halo Term}

At radii larger than about $3$ Mpc, contributions to the cluster density from other halos become important to model. To account for the 2-halo term, we model the total profile as
\begin{equation}
    \Delta \Sigma(r) = \Delta\Sigma_{1h}(r) + \twohaloA{} \Delta\Sigma_{2h}(r)
\end{equation}
where $\twohaloA{}$ is an arbitrary amplitude we marginalize over. We fix the masses in the 2-halo term to be the SZ-masses, rather than correcting them with $a_\obs{}$, in order to make $a_\obs{}$ independent of the absolute amplitude of $\Delta\Sigma_{2h}$. This is important because the amplitude is controlled by the halo bias $b(M)$ (see Equation \ref{eq:2h-corr} below), which is not known to high enough accuracy to justify coupling its predicted value into our model.

This summative approach contrasts with $\max(1h, 2h)$ models such as \citep{HayashiWhite2008}, where the 2-halo term only contributes at radii where it exceeds the 1-halo term. We do not take such an approach here as it produces poor results in the tests performed and shown in this work. Models with a $\max(1h, 2h)$ approach must cope with degenerate values of $\twohaloA{}$, as once this parameter becomes low enough it completely suppresses the 2-halo contribution, as do all other amplitudes less than this critical value.

To calculate $\Delta \Sigma_{2h}$, we use the same techniques from Appendix \ref{app:rho_to_delta_sigma_calc} to calculate the ESD from the density, and model the density as
\begin{equation}
    \rho_{2h}(r) = \rho_m(z=0) \xi_{2h}(r, M, z),
\end{equation}
where we subtract the constant background density component $\rho_m$. Note again that $r$ is a comoving radius and the matter density is the comoving matter density (hence evaluated at $z=0$). The 2-halo correlation function is
\begin{equation}
    \xi_{2h}(r, M, z) = b(M) \xi_{mm}(r, z), \label{eq:2h-corr}
\end{equation}
where $b(M)$ is the halo bias function \citep[e.g.,][]{cole-kaiser-1989, mo-white-1996, sheth-etal-2001, tinker-bias}, and $\xi_{mm}$ is the matter-matter correlation function obtained from the matter power spectrum:
\begin{equation}
    \xi_{mm}(r, z) = \frac{1}{2\pi^2} \int_{0}^{\infty} \dd{k} k^2 P_{mm}(k, z) j_0(kr),
\end{equation}
where $P_{mm}$ is the linear matter-matter power spectrum and $j_0$ is the zeroth spherical Bessel function of the first kind, $j_0(kr) = \sin(kr)/(kr)$. We make use of the MCFIT\footnote{\url{https://github.com/eelregit/mcfit}} software package to calculate $\xi(r)$ from $P(k)$.

We use the fitting function from \citep{tinker-bias} to estimate the linear bias, which is a function of $\Delta_m$. As we use $500c$ masses in this work, we convert $\Delta_c=500$ to the $\Delta_m$ value which corresponds to the same halo mass. Setting $M_{\Delta_m m} = M_{\Delta_c c}$ implies
\begin{align}
    \Delta_m V(R_{\Delta_m}) \rho_m(z) = \Delta_c V(R_{\Delta_c}) \rho_c(z),
\end{align}
yielding
\begin{equation}
    \Delta_m = \frac{\rho_c(z)}{\rho_m(z)} \Delta_c,
\end{equation}
which follows since the masses are the same and thus the volume $V(R_{\Delta_m})$ must equal $V(R_{\Delta_c})$. We then use the resulting $\Delta_m$s with the Tinker function, giving the desired $b(M_{500c})$.

\subsection{Stacking Procedure}

The basic premise of our model is that there is a galaxy cluster catalog of $N$ clusters (indexed by $i$), with observable-masses $\Mobs{}_i$ and redshifts $z_i$. The goal of our model is to fit for the average parameters in the $\Mobs{}$-$M$ relation. We accomplish this by fitting the stacked weak-lensing signal $\ev{\Delta \Sigma}$ (In fact, we fit a scaled version of this signal; details are described in the following section). The model has a corresponding modeled WL signal for each actual cluster WL signal. For this reason we refer to this technique as sample matching.

This is to contrast with the alternative technique of integrating over the entire support of the mass function in mass and redshift, obtaining information about the cluster catalog in question by convolving with the selection function of the survey. In our tests on simulated clusters, we have found that the latter produces a large bias in the inference of the $\Mobs{}$-$M$ parameters, which the matching stack technique is able to obtain nearly zero bias. We expect this is due to limitations in the simulation we test on not containing a fully representative sample of clusters, particularly rare high-mass clusters. The matching stack technique avoids depending on any assumptions about the cluster mass and redshift distributions.

To state this another way, instead of modeling cluster selection through a mass function and selection function, we directly obtain the information about the selected clusters by using the actual detected SZ-masses and cluster redshifts to model the stacked profile. We use an emulator of this matching model for the actual fitting process, which we describe in Section \ref{sec:emulator}. Below, we describe the details of this matching stack procedure.

\subsubsection{Matching Stack Model}

First we transform the weak-lensing ESD to a scaled quantity $D(r) = r \, \Delta \Sigma(r)$. This transformation emphasizes the effects of cluster parameters on the WL profile, and also compresses the range of values taken by the profiles (important for later emulating these profiles). A complete weak-lensing observation then constitutes the individual cluster weak lensing signals $D_i$, the cluster redshifts $z_i$, the lensing weights $w_i$, and the SZ masses $\Mobs{}_i$. Note that $w_i$ refers the normalized weights $\tilde w_i/\sum_j \tilde w_j$ where $\tilde w_i$ are the raw lensing weights. In our tests on simulated clusters, we set $w_i = 1$, but in the case of observed lensing data it is useful to use inverse-variance or similar weighting to enhance signal-to-noise.

The stacked weak-lensing signal is then
\begin{equation}
    \ev{D}(r) = \sum_i w_i D_i(r).
\end{equation}

The Matching Stack Model consists of evaluating an equal number of model $D$ profiles, evaluated at each of the $z_i$ and $M(\Mobs{}_i)$:
\begin{equation}
    \bar D(r) = \sum_i w_i D\qty(r, z_i, M(\Mobs{}_i; a_\obs{}), \vb*{\theta}), \label{eq:no-scatter-matching-model}
\end{equation}
where $D$ is a model of the lensing signal that depends on parameters $\vb*{\theta}$, which in the case of the full baryon model are $(c, \alpha, \beta, \gamma)$, and 
\begin{equation}
    M(\Mobs{}_i; a_\obs{}) = \frac{\Mobs{}_i}{\exp(a_\obs{})}. \label{eq:sz-mass-relation-noscatter}
\end{equation}

The above model ignores the scatter in $M$--$\Mobs{}$ relation, and mass dependent bias. While it is analytically simple to express scatter in the $M$--$\Mobs{}$ relation in our model, computationally this renders the calculation considerably more difficult. Adding a miscentering correction, which accounts for supression of the lensing signal by errors in the choice of cluster center, similarly is quite computationally expensive. We therefore do not use observable-mass scatter or miscentering corrections to obtain our results in Section \ref{sec:results}, which are based on Equation \ref{eq:no-scatter-matching-model}.

We have developed the formalism for mass-observable scatter in our stacking methodology, described in Appendix \ref{app:scatter}, and we are currently modifying our software implementation to overcome the computational difficulties, allowing both miscentering and scatter to be captured in this model.

\subsection{Summary}

The overall picture of our model is that (in the 0-scatter case) we assign each cluster a baryon profile-plus-dark matter only profile, with two-halo term correction, parameterized by 5 free parameters, then stack these model profiles together, weighting by the inverse-variance weights obtained from the WL signal. These profiles are evaluated at the redshift of the cluster; the mass is biased by the parameterized $M$-$\Mobs{}$ relation. In the case of nonzero $M$--$\Mobs{}$-scatter, each cluster's profile is integrated over a distribution of masses, whose integration measure is the mass function times the $\Mobs{}$ probability distribution. Selection is fully encapsulated by using $\Mobs{}$ and redshifts from the detected clusters, so there is no modeling of the selection function.

\section{Emulating the Weak Lensing Signal} \label{sec:emulator}

The stacking procedure and the corrections outlined in Section \ref{sec:wl} form a numerically complex model. While the calculation is not slow by the standard of evaluating the model for a single set of its parameters (generally $a_\obs{}, c, \alpha$, and $\beta$, while we fix $\gamma = 0.2$), fitting the model to stacked lensing data via MCMC would take days of computation time on most personal computers. In order to facilitate much faster inference, we construct an emulator (also known as a surrogate model) for the WL signal $\bar D(r, \vb*{\theta})$. 

The summary of this approach is that we decompose $\bar D(r, \vb*{\theta})$ using Principal Component Analysis (PCA) \citep{pearson-1901}, then interpolate the PCA weights over the parameters $\vb*{\theta}$ using a Gaussian Process interpolator. This approach is extremely similar to that taken in \citep{coyote-universe-emulator}. 

Unlike in \citep{coyote-universe-emulator}, our emulator is meant to be recalculated for each use. Because the stacked WL signal itself is unique to the sample of galaxy clusters selected, and our model reflects this, $D(r, \vb*{\theta})$ is a function of the clusters that are stacked. Thus, mass inference using our approach looks like this:
\begin{enumerate}
\item $\bar D(r, \vb*{\theta})$ is precalculated on the desired parameter space, using a sampling scheme described below
\item these samples are used to construct an emulator for a particular sample of cluster WL signals
\item the parameter space is resampled randomly and the emulator's accuracy tested at these new samples
\item the emulator is then used for MCMC fitting of the parameters
\end{enumerate}

Because the 1 and 2-halo contributions to $D(r)$ are summed, and the 2-halo term is fixed per cluster sample (i.e., we do not use 2-halo information), we do not need to emulate the $\twohaloA{}$ parameter but simply emulate the 1-halo parameters, and then use 
\begin{equation}
    \mathrm{emulator}_{1h}(\vb*{\theta}) + \twohaloA{} D_{2h}
\end{equation}
as the model for MCMC. In practice, we use the techniques described below to emulate the 2-halo term over mass and redshift for reasons of computational convenience, but the accuracy of this emulation is orders of magnitude better than that of the 1-halo emulator, which is already accurate to better than $1\%$.

In this section we detail how the emulators are constructed, and the sampling scheme used for probing the parameter space. In Appendix \ref{app:emulator-accuracy}, we detail tests of the emulator's accuracy. We perform these accuracy tests on every emulator used for each individual analysis performed in this work.

\subsection{PCA Emulator Construction}

As described previously, we emulate not $\Delta \Sigma(r)$ directly, but $D(r) = r \Delta \Sigma(r)$. This scaling compresses the data into profiles which vary over a narrower range of scales; both PCA and GP interpolation perform poorly on data that varies over multiple orders of magnitude.

\begin{figure}[!htbp]
    \centering
    \includegraphics[width=0.48\textwidth, keepaspectratio]{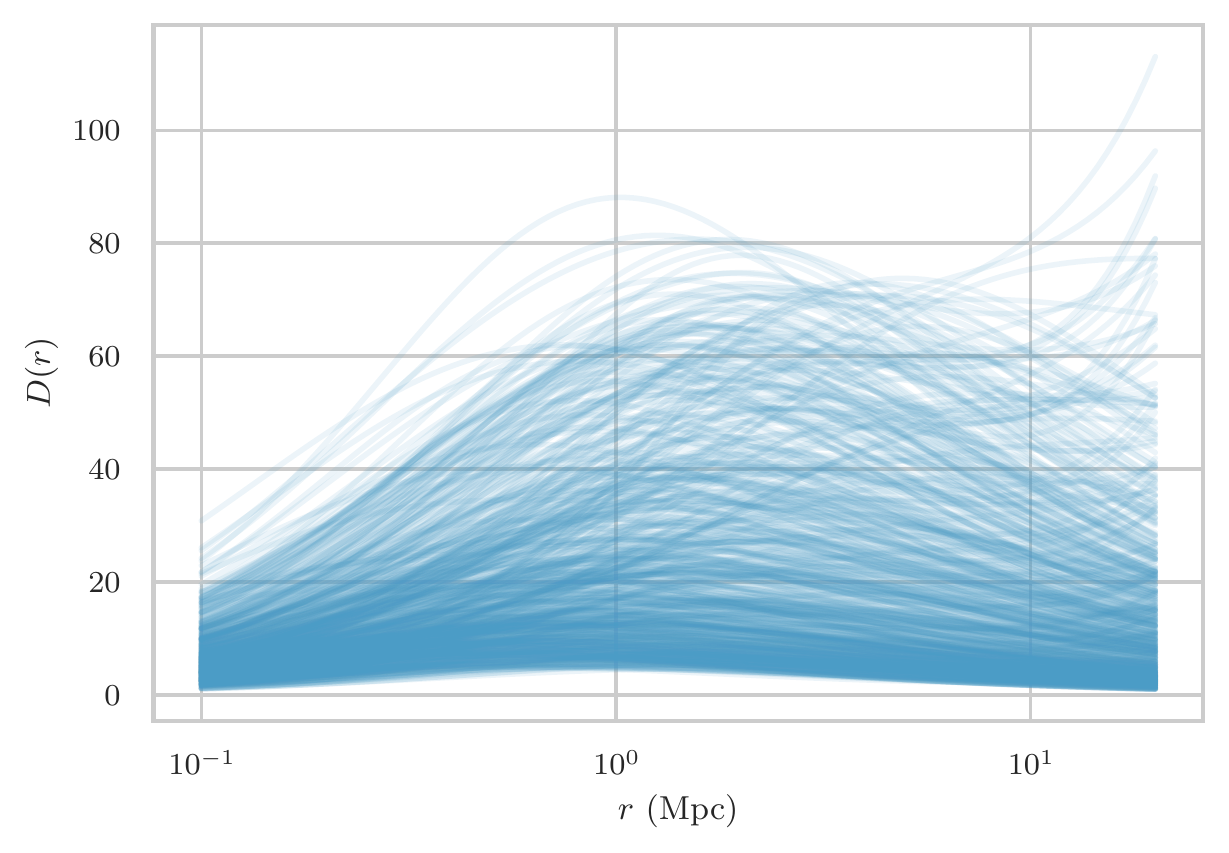} 
    \includegraphics[width=0.48\textwidth, keepaspectratio]{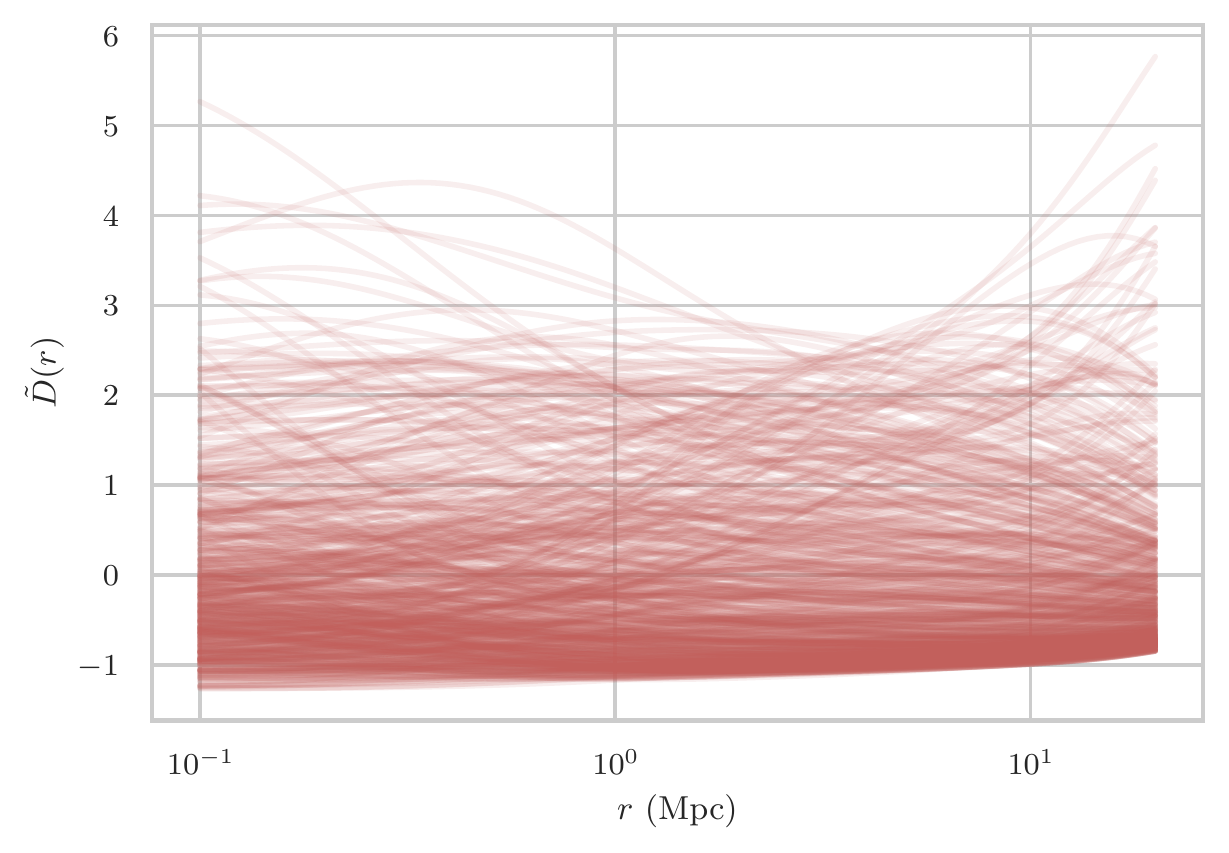}
    \caption{\label{fig:lh-wl-signal} Left: lensing signal $\bar D(r, \vb*{\theta})$ curves of different parameter combinations on a Latin hypercube of size 600 $\vb*{\theta}$ samples. Right: the scaled and shifted lensing signals $\tilde D(r, \vb*{\theta})$ of the same sample. The regularization of the sampled data prepares it for principal component analysis.}
\end{figure}

Before emulation the WL signal $\bar D(r, \vb*{\theta})$ is scaled by having its mean over the parameter samples subtracted, and being divided by its standard deviation over the parameter samples:
\begin{equation}
    \tilde D(r, \vb*{\theta}) = \qty[\bar D(r, \vb*{\theta}) - \ev{\bar D(r)}_{\vb*{\theta}}] 
    \big / \mathrm{std dev}(\bar D(r))_{\vb*{\theta}}
\end{equation}
Examples of pre- and post-scaling WL signals are shown in Figure \ref{fig:lh-wl-signal}.

\subsubsection{PCA Decomposition}

The emulator is constructed from the scaled WL signal using two steps; first, $\tilde D(r, \vb*{\theta})$ is decomposed into a hierarchical basis of functions using PCA; second, the coefficients of these basis functions are interpolated over the parameter space.

We first tackle the decomposition itself. The desired form of this decomposition is
\begin{equation}
    \tilde D(r, \vb*{\theta}) = \sum_i w_i(\vb*{\theta}) \phi_i(r), \label{eq:wl-decomp}
\end{equation}
where explicitly the coefficients $w_i(\vb*{\theta})$ depend only on the parameters $\vb*{\theta}$ and the basis functions $\phi_i(r)$ only on radius. We only then need to interpolate $w_i$, which decouples the radial dimension from interpolation altogether. This results in much lower error levels than directly interpolating the entire function $\tilde D(r, \vb*{\theta})$ over $\vb*{\theta}$.

In order to construct the $w_i$ and $\phi_i$, we use PCA. PCA can be performed using Singular Value Decomposition (SVD). Suppose that we represent $\tilde D(r, \vb*{\theta})$ as a $(N_r, N_s)$ matrix $\mathcal{D}$, with $N_r$ the number of radial bins, and $N_s$ the number of samples over $\vb*{\theta}$. Then the SVD decomposition of this matrix yields
\begin{equation}
\mathcal{D} = U S V^\intercal,
\end{equation}
where $U$ is an orthogonal $(N_r, N_r)$ matrix, $S$ is a diagonal $(N_r, N_r)$ matrix whose diagonal values are the singular values of $\mathcal{D}$, and $V$ is an orthonormal $(N_r, N_s)$ matrix. In the context of PCA, the principal components (PCs) are the columns of $V$, and rows of the matrix $US$ contain the weights on these PCs.

\begin{figure}[!htbp]
    \centering
    \includegraphics[width=\textwidth, keepaspectratio]{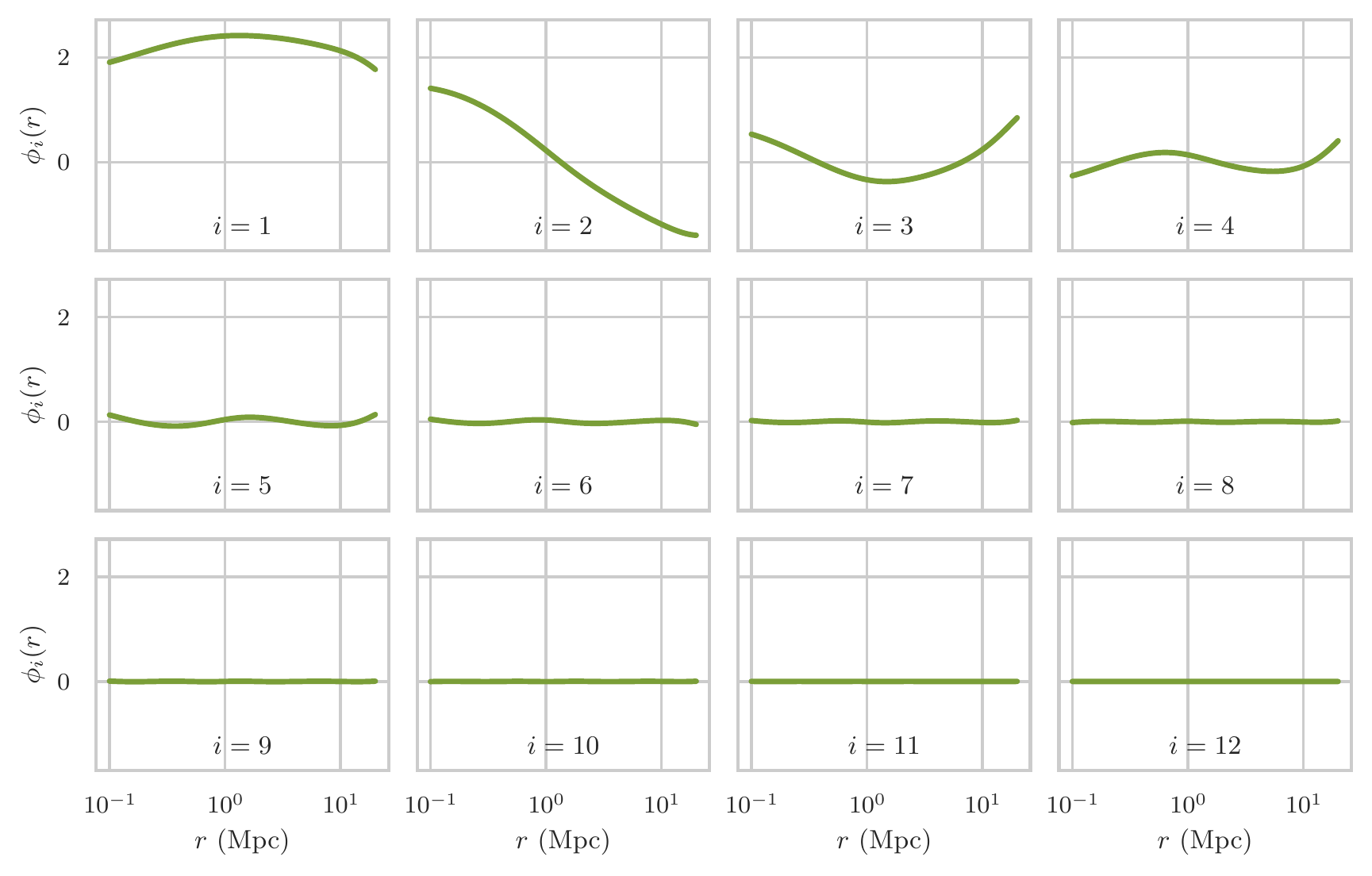}
    \caption{\label{fig:12pcs} The first 12 PCA basis vectors $\phi_i$ for a sample of 600 stacked $\tilde D(r, \vb*{\theta})$ profiles. The diminishing impact of the basis vectors can be seen reflected in the decreasing amplitude with each $\phi_i$.}
\end{figure}

However in the context of our decomposition in Equation \ref{eq:wl-decomp}, the basis vectors $\phi_i$ are proportional to the rows of $US$ and the weights $w_i$ the columns of $V$. An example of the first 12 of these basis vectors is shown in Figure \ref{fig:12pcs}. We choose a scaling factor $\sqrt{N_r}$ to scale both the weights and vectors, so that
\begin{align}
    \phi_{ir} &= (US)_{ir} / \sqrt{N_r}, \\
    w_{i\theta} &= \sqrt{N_r} V_{i\theta}.
\end{align}

\begin{figure}[htbp!]
    \centering
    \includegraphics[width=0.5\textwidth, keepaspectratio]{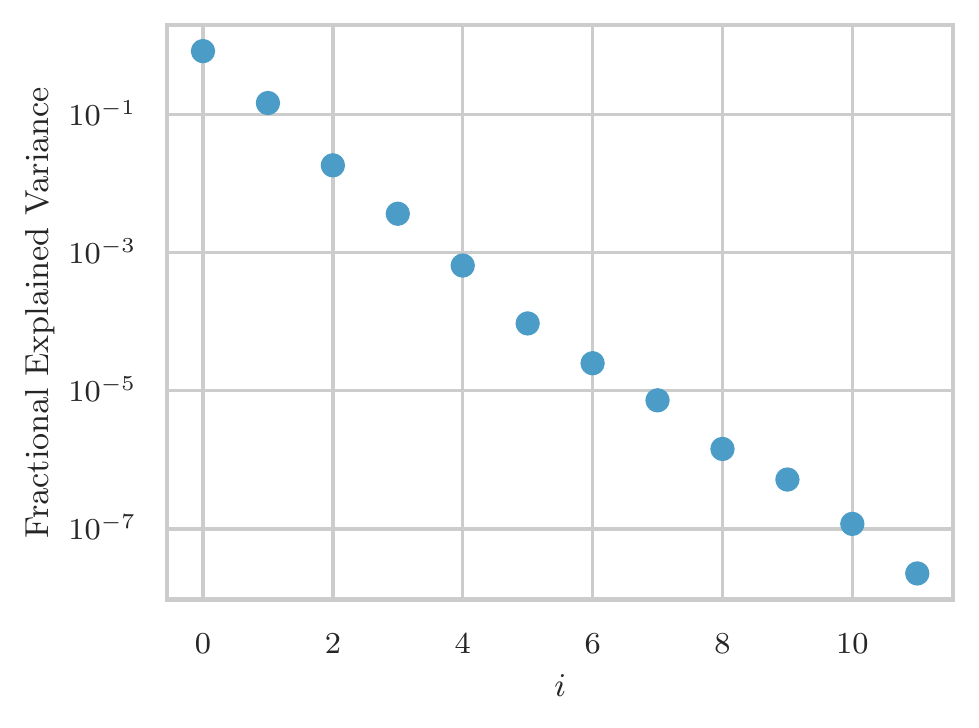}
    \caption{\label{fig:explained-variance} Fractional variance accounted for by each principal component of a sample stacked WL signal, plotted as a function of the principal component index $i$. The decrease of this variance is close to exponential, allowing only a few principal components to be used for emulating the lensing model.}
\end{figure}

The primary advantage of PCA being used for this decomposition is that it gives a \emph{hierarchical} set of basis vectors; that is, each PC contributes a decreasing amount of variance to the data. Typically this decrease is exponential (see Figure \ref{fig:explained-variance}), and a small number of PCs can be used to adequately approximate most data. In this case, only about 8-10 PCs are needed to achieve highly accurate interpolations over 4 parameters (see Figures \ref{fig:errors-by-samps-and-comps} and \ref{fig:error-hists-by-components}).

\subsubsection{Interpolation}

Once the PCA decomposition is completed, the next step is to interpolate the weights $w_i(\vb*{\theta})$ over the parameter space. There are several interpolation techniques which are viable choices for multidimensional, sparsely sampled data, however the one which we find performs best is Gaussian Process (GP) interpolation, also known as Gaussian Process Regression or kriging. 
A detailed discussion of GP interpolation is beyond the scope of this work, but in short, the data points being interpolated are modeled as correlated Gaussian noise, with a specified covariance function \citep[e.g.][]{rasmussen-williams-gp}. The input samples are taken as an "observation", and then the interpolant is the mean of the conditional distribution of the GP, given that observation. Free parameters of the covariance function are determined through optimizing the likelihood function of the GP. 

The correlation function we use is the relatively common Matérn covariance function \citep[described in, e.g.,][]{rasmussen-williams-gp}, which we find gives high accuracy interpolations. We use the Scikit Learn \citep{scikit-learn} implementation\footnote{\url{https://scikit-learn.org/stable/modules/gaussian_process.html}} of Gaussian Process interpolation.

\subsection{Parameter Sampling}

One problem with GP interpolation (and indeed other methods used for greater than two-dimensional interpolation) is that it requires a square $(N_s, N_s)$ matrix be created where $N_s$ is the number of samples to be interpolated. This means that the computer memory usage of this interpolation scales like $N_s^2$.

The usual approach to improving interpolation accuracy is to increase the number of samples. However, in our case (a $4$-dimensional parameter space), the density of uniformly spaced samples grows like $N_s^{1/4}$. Thus it is prohibitively difficult to uniformly sample the parameter space finely enough for high accuracy, while staying within computer memory limits.

One mitigation for this problem is not to use uniform sampling. An obvious choice is uniform Monte-Carlo sampling, where points are randomly chosen within the given parameter space. This choice is better than a uniform grid, but it suffers due to clustering. This leaves large voids where there are few samples, increasing the interpolator error in that void.

A better solution is to use Latin Hypercube (LH) sampling \citep[e.g.,][]{mckay-etal-lh, sacks-etal-lh, currin-etal-lh}. The most straightforward definition of a LH is through generalization of its definition for 2 dimensions. In 2D, a LH is a collection of points arranged on a square grid, such that each row of the grid and each column of the grid contain exactly 1 point.

Formally, let $d$ be the dimension of a hypercube, and let this hypercube be divided into $N^d$ hypercubic cells. Then, index each cell by a $d$-tuple $(n_1, n_2, \dots, n_d)$, where $n_i$ numbers the position of the cell along a the $i$th dimension of the hypercube. Then, any $d$ unique permutations of $(1, 2, \dots N)$ together form the cell-coordinates of a Latin Hypercube of size $N$.

\begin{figure}[!htbp]
    \centering
    \includegraphics[width=0.48\textwidth, keepaspectratio]{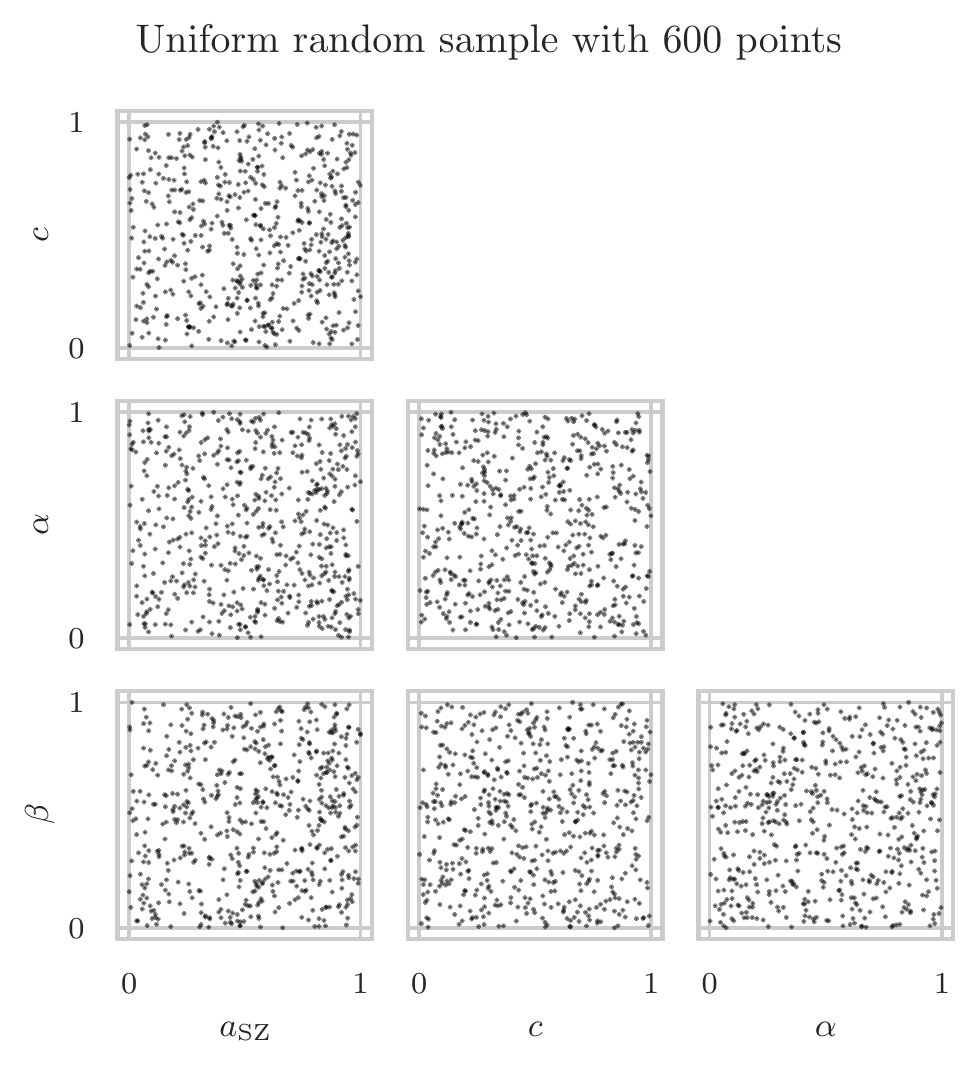}
    \includegraphics[width=0.48\textwidth, keepaspectratio]{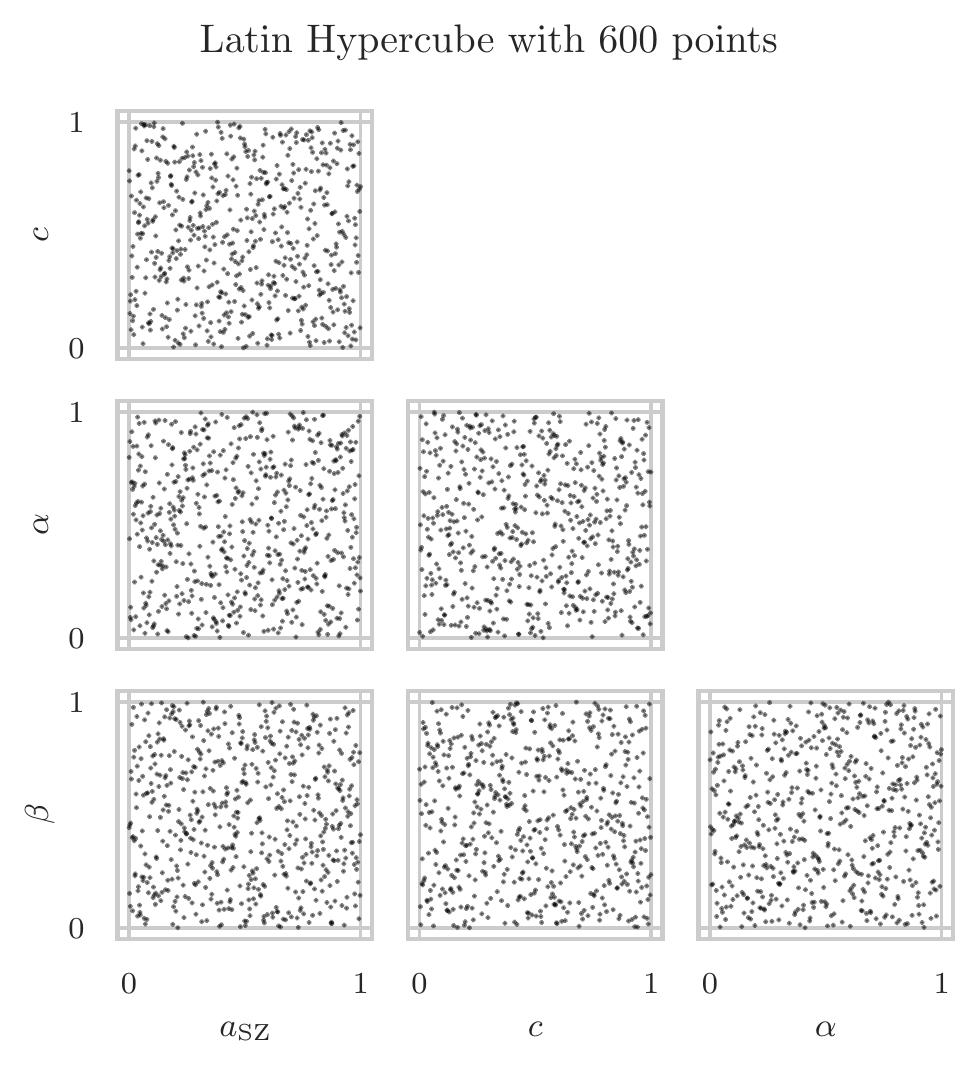}
    \caption{\label{fig:lh-vs-rand} Left: A uniform random sample of parameters in 4 dimensions, projected to 2D planes for every pair of axes, to demonstrate the level of clustering at this sample size. Right: Same but for a Latin Hypercube sample. Note that the parameter axes are scaled and shifted such that all parameters run from 0 to 1.}
\end{figure}

LH sampling has the advantage of reducing clustering while being arbitrarily sparse. In Figures \ref{fig:lh-vs-rand} we show an example 4-dimensional uniform Monte-Carlo sample and an LH sample for comparison.

We generate LH samples by randomly selecting permutations of integers, then stretching and shifting each dimension of the LH to match the desired limits of the actual parameters in question. There are techniques of refining LH samples to further spread out points and achieve a more optimal sample, however for sample sizes greater than $\sim 100$ we find random LH samples are sufficient (see results of Appendix \ref{app:emulator-accuracy}).

\section{Results}
\begin{figure}[!htbp]
    \centering
    \includegraphics[width=0.49\textwidth, keepaspectratio]{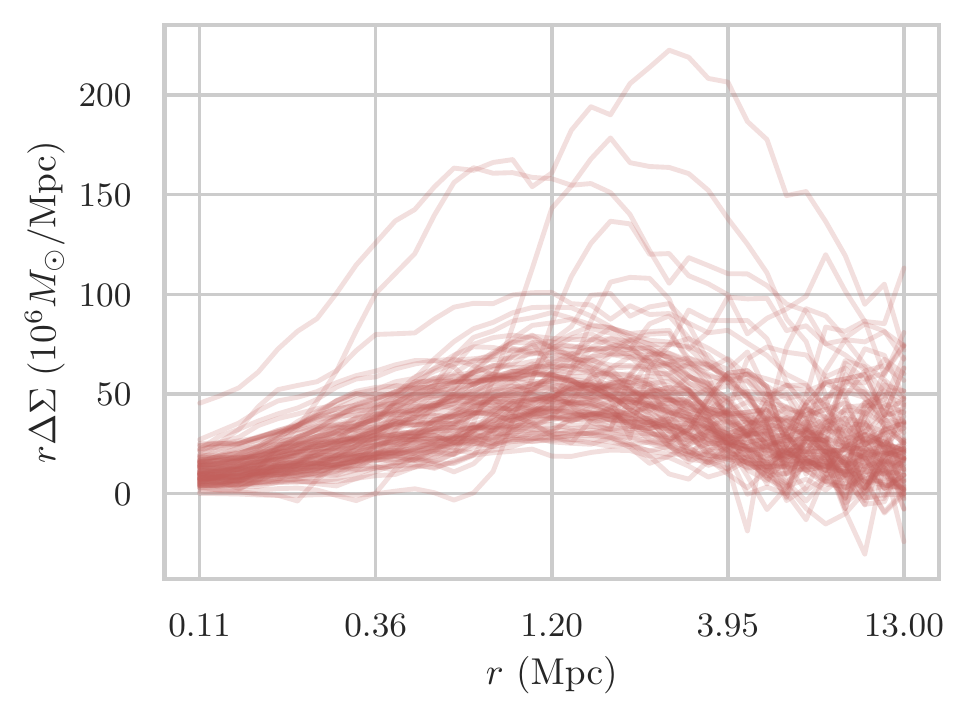}
    \includegraphics[width=0.49\textwidth, keepaspectratio]{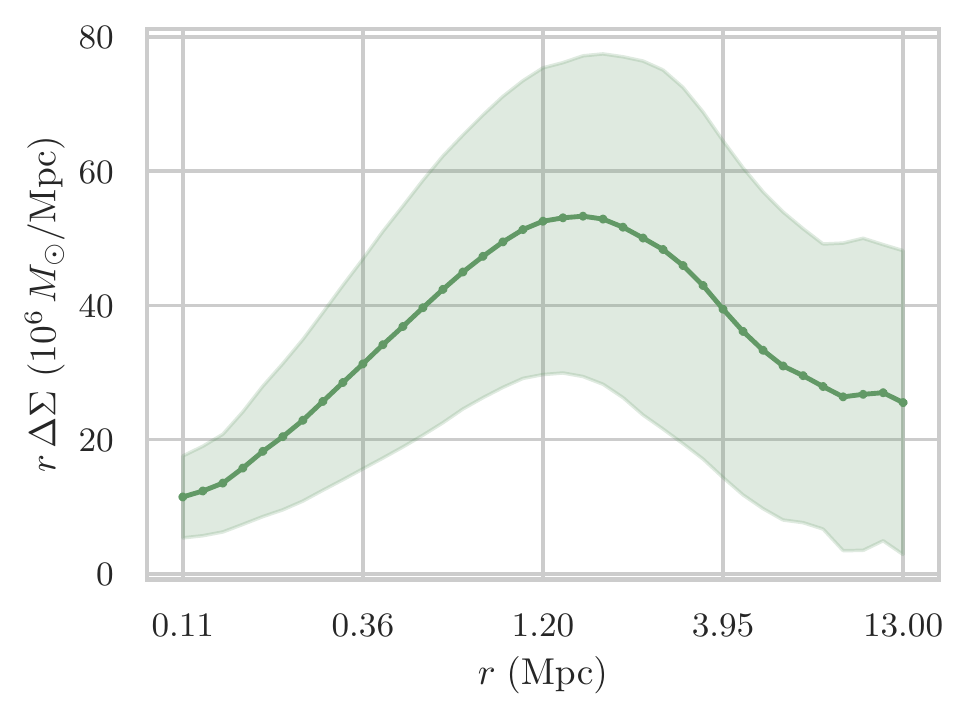}
    \caption{\label{fig:sims} Left: 100 randomly selected cluster $D(r)$ profiles out of the total sample of 1400. Only 100 profiles are shown to increase visibility in the plot. We note that we cut the inner radii below $0.1$ Mpc to avoid unphysical noise due to finite simulation resolution. Right: The stacked simulation weak-lensing signal $D(r)$, along with the 1-$\sigma$ region of the 1400 profiles.}
\end{figure}

Using simulations, we validate that our model indeed reduces the bias in fitting the relation between SZ mass and WL mass. The simulations used for excess surface density profiles are obtained from the simulated galaxy clusters presented in \citep{battaglia-simulations}. We use the top 100 most massive clusters in each of $14$ redshift bins. A sample of 100 randomly selected cluster WL signals out of these 1400 are shown in the left panel of Figure \ref{fig:sims}. Using simulations rather than real weak lensing data allows a controlled test, where we know the true cluster masses and many extraneous corrections can be ignored. This allows us to test specifically whether our model reduces bias relative to a model without baryons. To create the stack, we simply average these 1400 profiles together; this is shown in the right panel of Figure \ref{fig:sims}. Thus, our stack uses weights of unity for all clusters.

We run MCMC fits of both our baryonic model, and a version of our model that only includes the CDM (NFW) term to test whether our model is less biased than not using any baryonic density term. We use the actual simulated cluster masses as the SZ-masses; this means that if our model is unbiased, the posterior of $a_\obs{}$ should be consistent with $a_\obs{} = 0$. This test is also consistent with not including any scatter in the SZ-mass relation, as discussed earlier in Section \ref{sec:wl}.

\subsection{Likelihood and Covariance} \label{sec:likelihood}

Because we fit a stacked profile, we can use a Gaussian likelihood function for the tests on simulations:
\begin{equation}
    L(\vb*{\theta}) = \frac{1}{\sqrt{(2\pi)^d \det{\mathcal{C}}}} \exp(-\frac{1}{2} \qty[\bar D(r, \vb*{\theta}) - D_{\mathrm{sim}}(r)]^\intercal \mathcal{C}^{-1} \qty[\bar D(r, \vb*{\theta}) - D_{\mathrm{sim}}(r)]),
\end{equation}
where $D_\mathrm{sim}(r)$ is the stacked simulation WL signal. The covariance matrix $\mathcal{C}$ is a transformation of the $\Delta \Sigma$ covariance matrix $C$, given by
\begin{align}
    \mathcal{C} = \vb*{r}^\intercal C \vb*{r},
\end{align}
where $\vb*{r}$ is the vector of radii used, in order to obtain the covariance of $D(r)$.
The ESD covariance matrix $C$ is estimated using the methods also used in \citep{miyatake-wl-2019}, where $C$ is constructed from a statistical shape-noise term $C^\mathrm{stat}$, and a large-scale-structure term $C^\mathrm{lss}$:
\begin{equation}
    C = C^\mathrm{stat} + C^\mathrm{lss}.
\end{equation}

\begin{figure}[!htbp]
    \centering
    \includegraphics[width=0.6\textwidth, keepaspectratio]{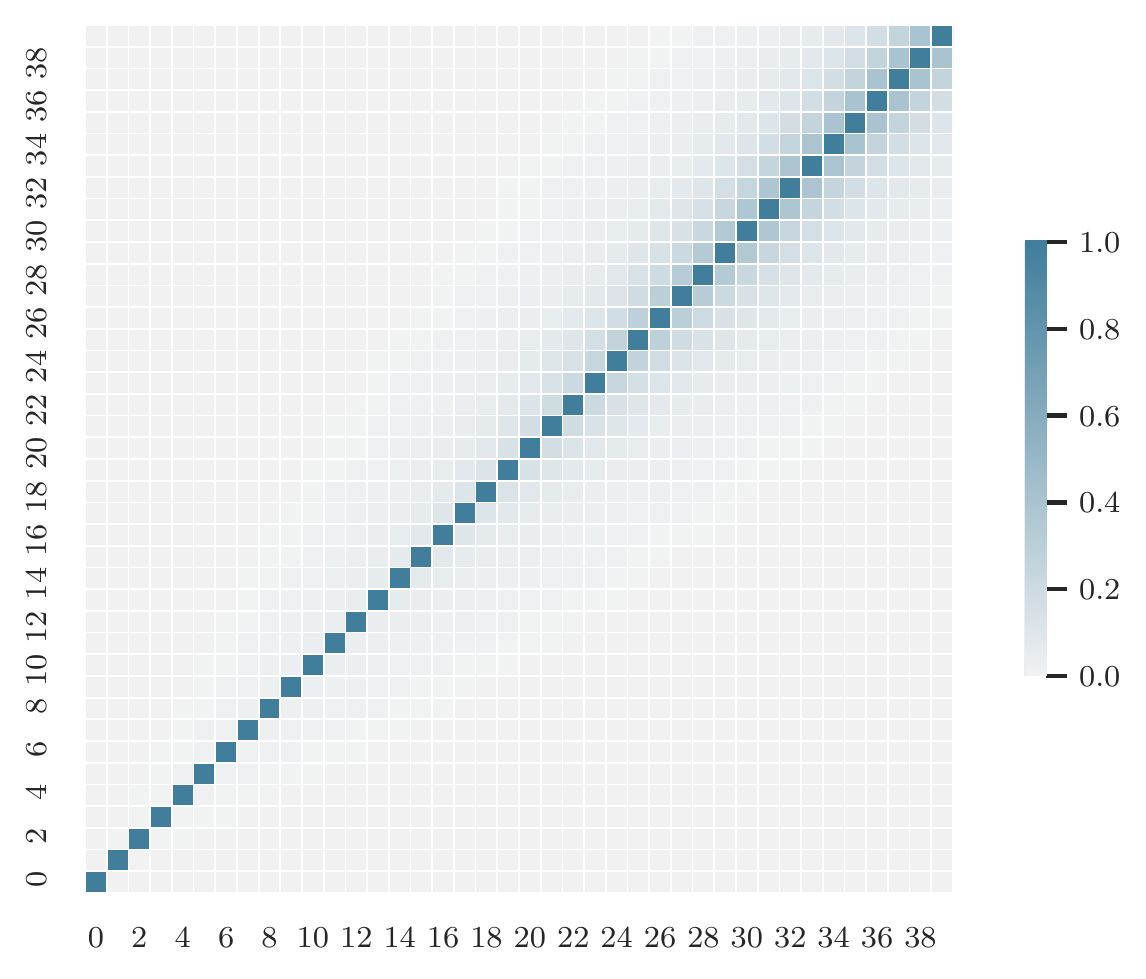}
    \caption{\label{fig:corr} Correlation matrix for the simulated cluster stacked excess surface density profile, indexed by radial bin number. The radial bins are uniformly log-spaced between $0.1$ Mpc and $12$ Mpc. The outer radii are more correlated than the inner radii.}
\end{figure}

To compute the covariance, we use a Hyper-Suprime Camera-like lensing source distribution, with source number density of $n_s=20$ galaxies per square-arcminute, with redshift distribution as used in \citep{oguri-takada-2011}. The full details of the calculation of these covariance matrices can be found in \citep{miyatake-wl-2019}. One difference our work has with \citep{miyatake-wl-2019} is we do not include a term for halo-triaxiality or correlated halos, as this term is sub-dominant at all scales \citep[see Figure 4 of][]{miyatake-wl-2019}. 
We show the resulting correlation matrix for our simulated clusters in Figure \ref{fig:corr}.

We note that this covariance matrix is designed to be representative of the current state-of-the-art in lensing measurements, but it contains galaxy shape-noise not present in the simulated data, so the variations in the simulated stacked profile are much smaller than what would be inferred from this covariance matrix.

\subsection{Comparing the Baryon and NFW-only models} \label{sec:results}

We perform tests over two sets of radial ranges to validate our model; one over the radial range $0.1, 12$ Mpc, where we include the two-halo term, and one over the range $0.1, 5$ Mpc, where we only include the 1-halo term. For each of these we fit an NFW-only model and a model including our baryonic correction, but the stacking procedure and all other calculations are the same. We use the second radial range to determine that the 2-halo component of the model and data is not the primary source of bias in the NFW-only model.

\begin{figure}[!htbp]
    \centering
    \includegraphics[width=0.98\textwidth, keepaspectratio]{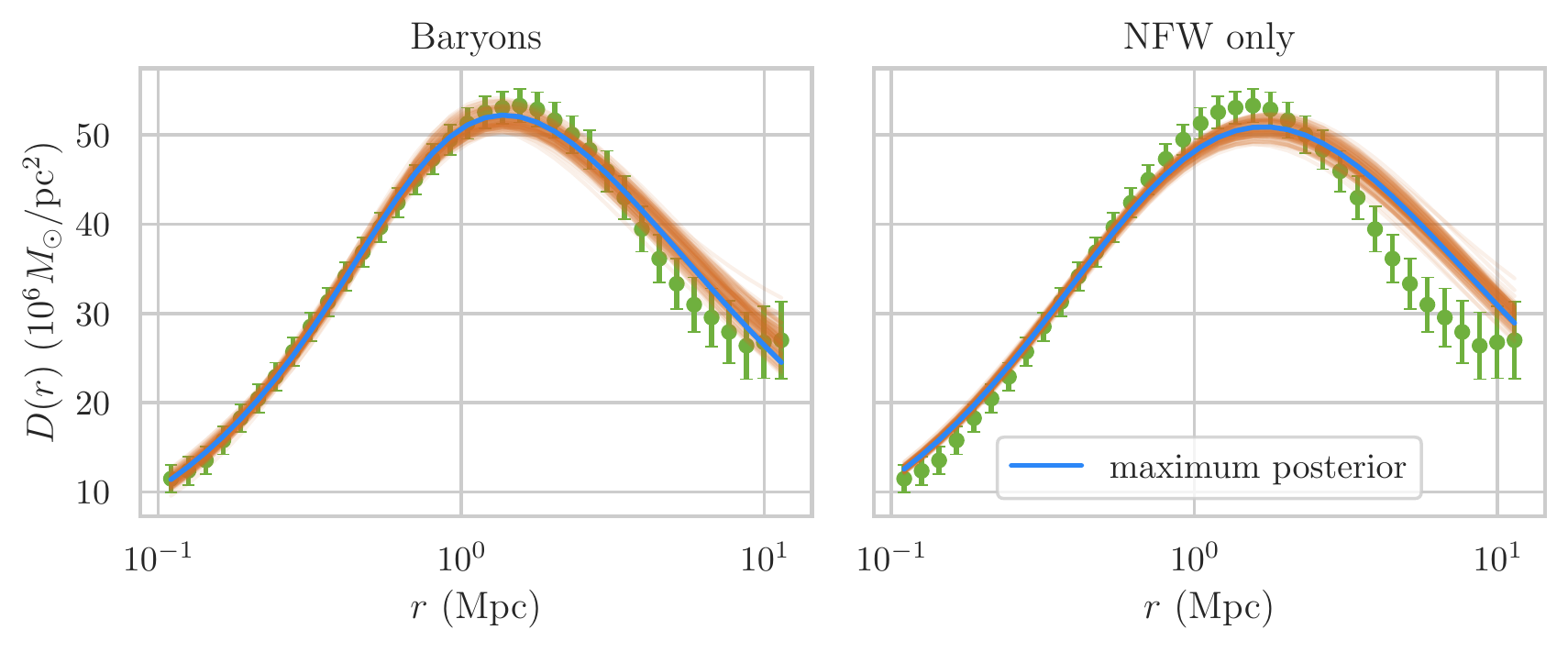} 
    \caption{\label{fig:2h-fits} Left: Maximum-probability baryon 1-halo + 2-halo model WL profiles (blue) shown with 100 profiles selected randomly from the $68\%$ confidence region of the posterior for reference (orange), and the simulation stack (green points with error bars). Right: same, for NFW-only 1-halo + 2-halo model. The inclusion of baryons produces a significantly better fit. Note that the high-radius bins are highly correlated, resulting in the systematic overestimation of $D(r)$ at most of these radii.}
\end{figure}

\begin{figure}[!htbp]
    \centering
    \includegraphics[width=0.49\textwidth, keepaspectratio]{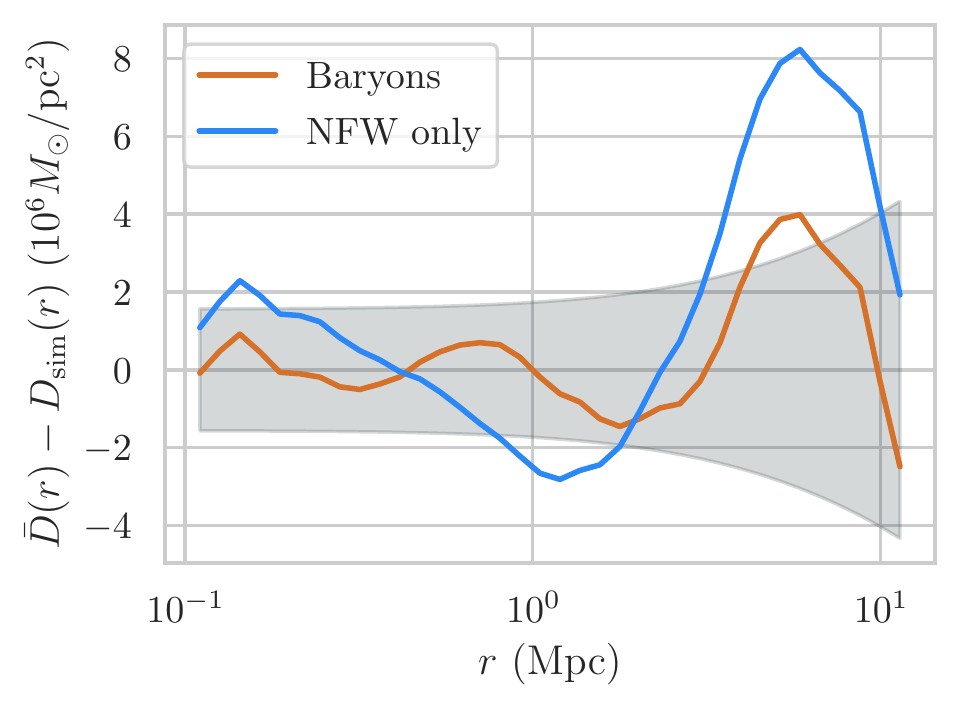}
    \includegraphics[width=0.49\textwidth, keepaspectratio]{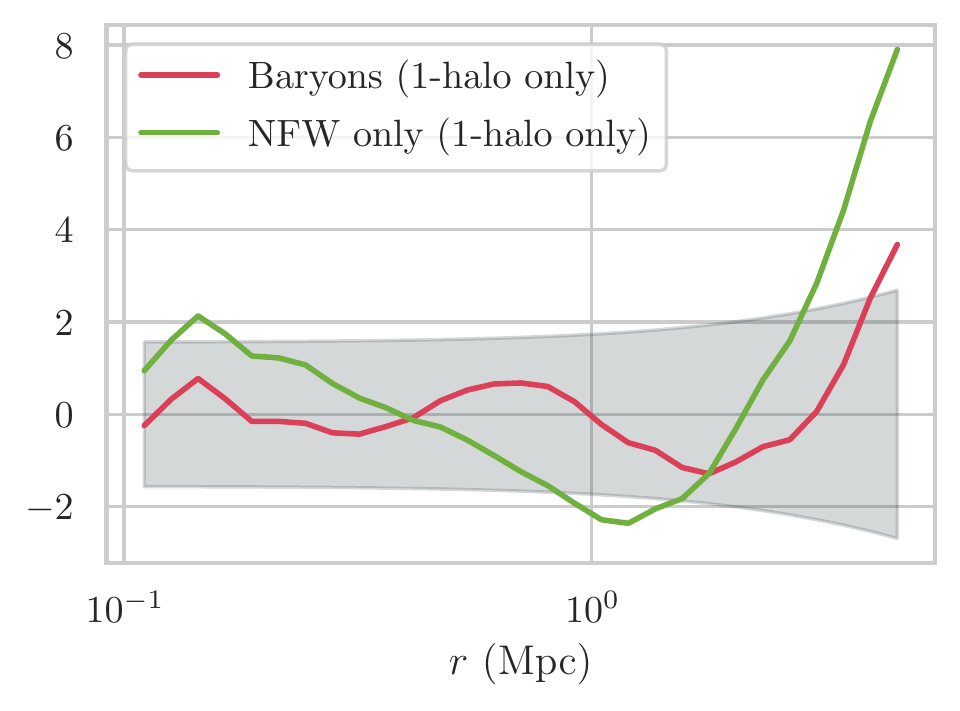}
    \caption{\label{fig:residuals} Left: Residuals for the peak-posterior profiles of the baryon model (orange) and NFW-only model (blue). Errors on the data are shown as a grey shaded region. The goodness of the baryon model's fit is superior, but we do not calculate $\chi^2$ or PTE values, as these values are uninformative due to the simulated data not containing the intrinisic variance captured in the weak-lensing covariance matrix. Right: Same, but over the restricted $0, 5$ Mpc radial range, with the one-halo only baryon (magenta) and one-halo only NFW (green) residuals, with the errors again shown by the grey shaded region. The baryon fit still maintains smaller residuals. Note in both cases that the residuals are correlated at large radii.}
\end{figure}

\begin{figure}[!htbp]
    \centering
    \includegraphics[width=0.98\textwidth, keepaspectratio]{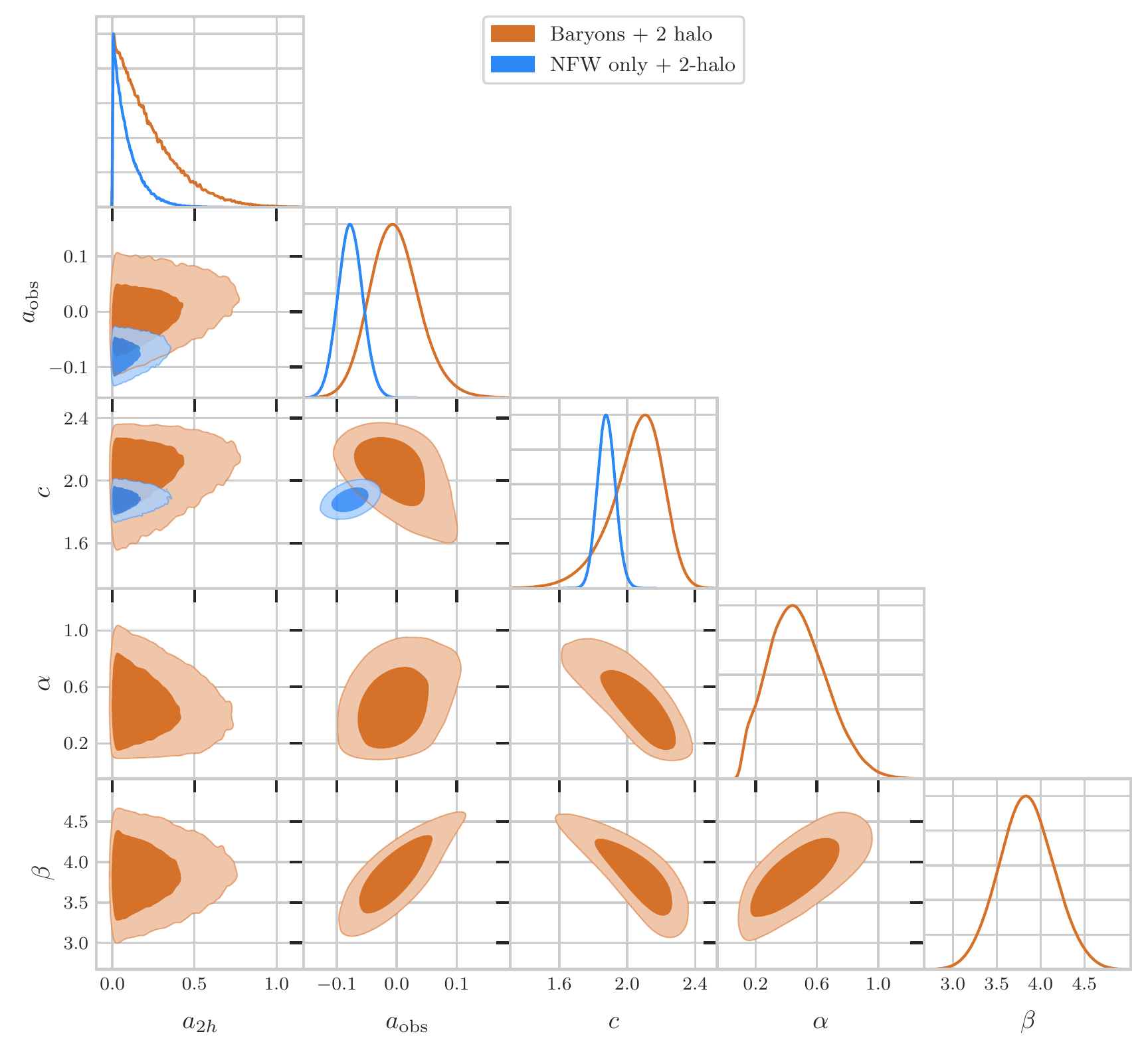}
    \caption{\label{fig:2h-corner} One-and-two-dimensional marginalized posteriors of the baryon (orange) and NFW-only (blue) 1-halo + 2-halo model parameters. Note that $a_\obs{} = 0$ is the expected result with an unbiased mass estimation; the baryon model posterior is clearly consistent with this value, the NFW-only posterior is not.}
\end{figure}

For the first set of results ($0.1, 12$ Mpc), we find that the baryon model is able to much more accurately fit the simulation stack than the NFW-only model, as seen in Figure \ref{fig:2h-fits} and the left panel of Figure \ref{fig:residuals}. The resulting fit finds a maximum-probability $a_\obs{}$ of $\fullRadiiBaryResult{}$ (which implies a percent bias on the mass of $\fullRadiiPctBiasBary{}$) when including baryons, and $a_\obs{} = \fullRadiiNfwResult{}$ ($\fullRadiiPctBiasNfw{}$) when only using an NFW density (marginalized posteriors of all parameters shown in Figure \ref{fig:2h-corner}).

There are several interesting features of these results: first, $\twohaloA{}$ has a peak-posterior value of essentially $0$. This is likely due to the errors being large at those radii, the lack of very many radial bins where the 2-halo term is relevant. This is particularly acute in the NFW-only case, because the fitted NFW model overpredicts the lensing signal significantly at those radii. The constraints for $c$ are very tight for NFW-only, but rather broad for the baryon model due to covariance with $\alpha$ and $\beta$. Both $\alpha$ and $\beta$ are weakly constrained, but have peaks roughly consistent with the results from \citep{tau-of-clusters}. Lastly and most importantly, the $a_\obs{}$ peak-posterior when including baryons is $\fullRadiiBiasNumSigmaAwayBary{}\sigma$ from $0$, but the NFW-only model's peak is $\fullRadiiBiasNumSigmaAwayNfw{}\sigma$ from $0$.

One feature of these results on the $0.1$--$12$ Mpc range is how the NFW-only model overpredicts the high-radii values by much more than the expected error at these radii. This brings forth the question as to whether the NFW-only model might perform better if not fit to such a large radial range. Thus, we performed the second test on the smaller range $0.1$--$5$ Mpc, where the 2-halo term is much less significant for most clusters. In this second set of results, we don't include any 2-halo term and fit over a smaller radial range. The $a_\obs{}$ values are $\cutRadiiBaryOneHaloOnlyResult{}$ ($\cutRadiiPctBiasBaryOneHalo{}$) with baryons and $\cutRadiiNfwOneHaloOnlyResult{}$ ($\cutRadiiPctBiasNfwOneHalo{}$) with an NFW-only model. The baryon $a_\obs{}$ value here is $\cutRadiiBiasNumSigmaAwayBaryOneHalo{}\sigma$ from $0$, and the NFW-only model's is $\cutRadiiBiasNumSigmaAwayNfwOneHalo{}\sigma$. The baryon fit is still superior to the NFW-only fit (see right panel of Figure \ref{fig:residuals}). We find little difference from the results including the 2-halo term, except that the peak bias parameters are both a bit further from $0$. This indicates that regardless of the 2-halo contribution, the NFW-only model still biases the estimate of the masses at roughly the same level, and the particularly poor quality of the NFW-only fit at high-radii is not the reason for this bias.

\begin{figure}[!htbp]
    \centering
    \includegraphics[width=0.98\textwidth, keepaspectratio]{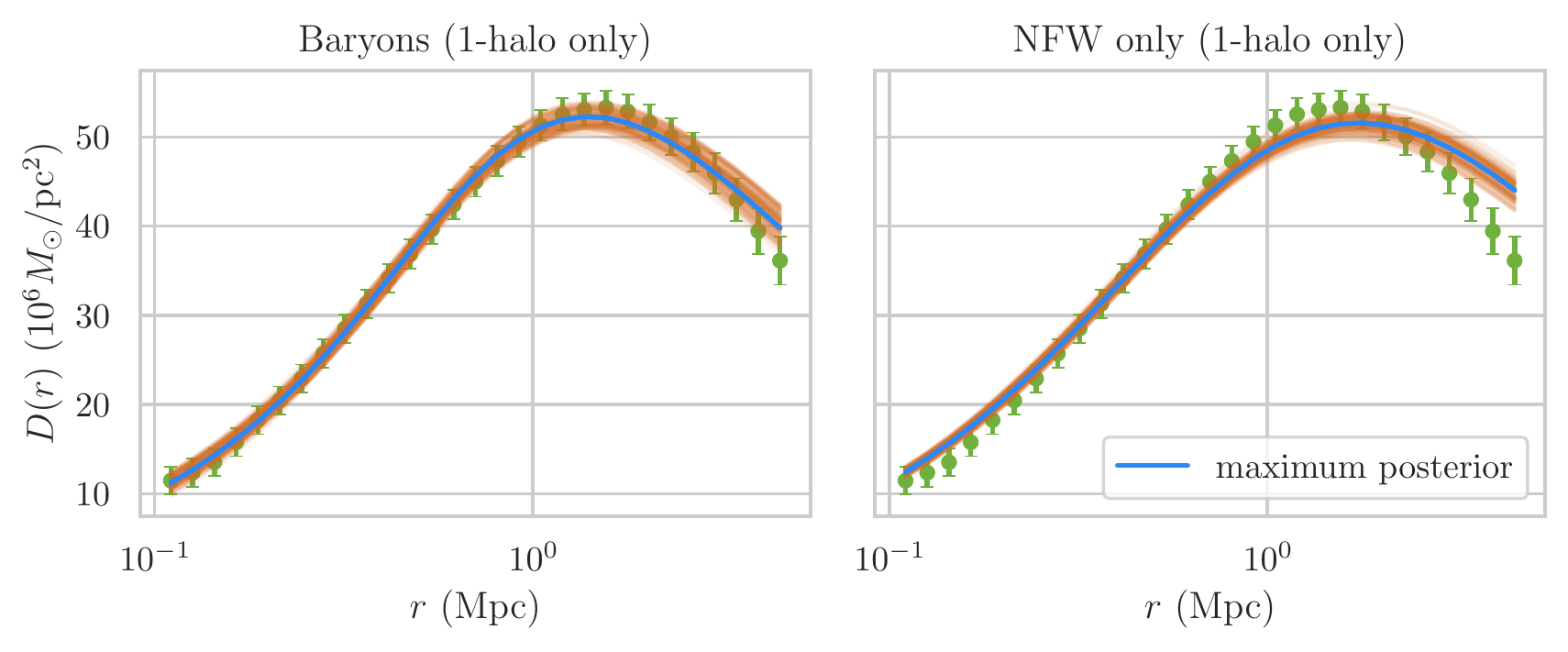}
    \caption{\label{fig:1h-fits} Same as Figure \ref{fig:2h-fits} but with the restricted $0$--$5$ Mpc radius range and no 2-halo term included in the model. While the NFW model performs somewhat better than over the full radii range, the baryon fit is still superior and the NFW model cannot accurately match the inner profile shape.}
\end{figure}

\begin{figure}[!htbp]
    \centering
    \includegraphics[width=0.8\textwidth, keepaspectratio]{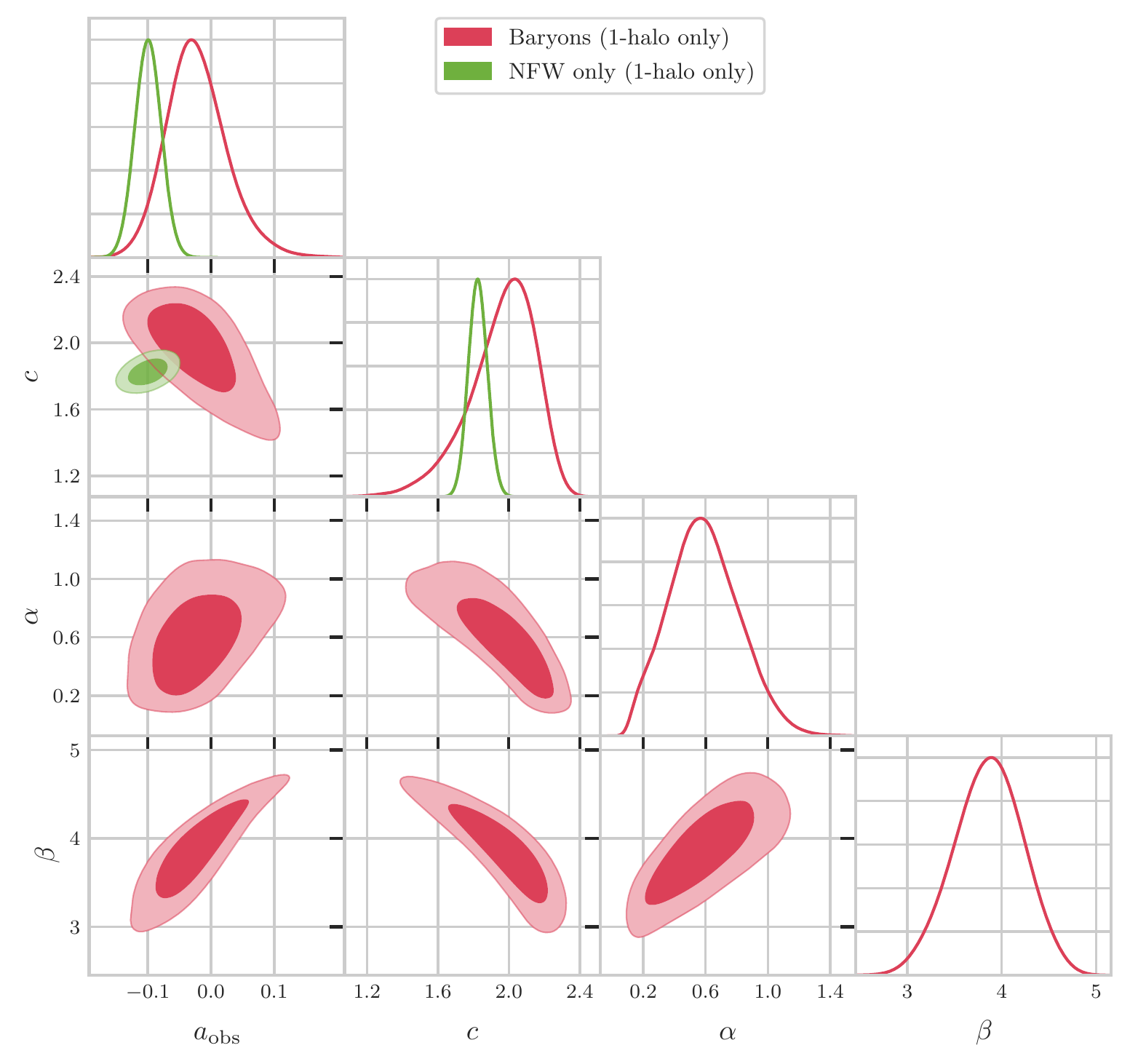}
    \caption{\label{fig:1h-corner} Same as Figure \ref{fig:2h-corner} but with the restricted $0$--$5$ Mpc radius range and no 2-halo term included in the model. The baryon model's peak $a_\obs{}$ value is further from 0 than when using the full radius range, but the NFW-only model also becomes more biased and is significantly more so than when including baryons.}
\end{figure}

\begin{figure}[!htbp]
    \centering
    \includegraphics[
    width=0.6\textwidth, 
    keepaspectratio,
    ]{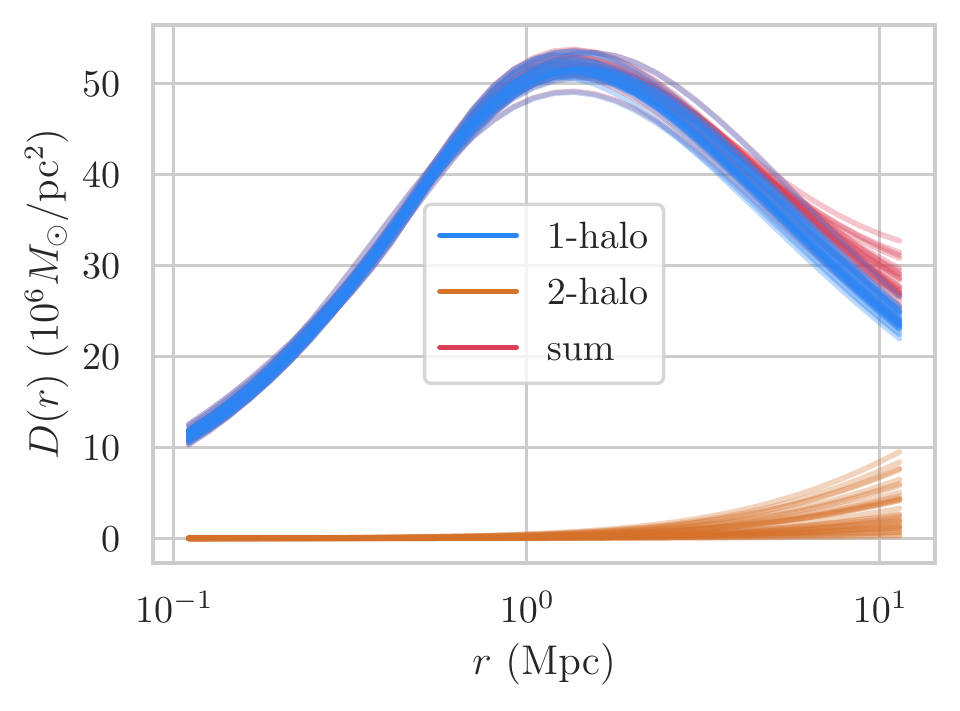}
    \caption{\label{fig:2halo-example} Stacked $D(r)$ profiles shown from the $68\%$ confidence region of the baryon model posterior, for the radius range $0.1$--$12$ Mpc (red), with the constituent 1-halo components (blue) and 2-halo components (orange) of each profile. The contribution of the 2-halo term is low, with the peak-posterior value being consistent with 0.}
\end{figure}

We illustrate the 2-halo contribution to the model, by showing 30 profiles from the $68\%$ confidence region of the $0.1$--$12$ Mpc baryon model results in Figure \ref{fig:2halo-example}. In this figure are shown the full profiles, in addition to the 1-halo and 2-halo terms individually. As seen in the figure, the 2-halo term contributes very little to the overall profile in the high-confidence region of the posterior, with few profiles having large 2-halo amplitudes. The small impact of the 2-halo term on this data partly explains why the $\mathrm{max}(1h, 2h)$ model performs worse than the additive model we employ here; under the $\mathrm{max}(1h, 2h)$ mode only fairly large 2-halo terms can contribute at all to the total profile. This would likely change if the simulated data had large 2-halo terms, or if the simulated data extended to a larger radial range, past 12 Mpc.

Finally, we note that we actually tested all 4 models (Baryon 2h, NFW 2h, Baryon 1h, NFW 1h) on both radial ranges as a sanity check. The results not described previously (1-halo only on the larger radial range, 1-halo + 2-halo on the restricted range) were not different by much, except that the 2-halo models fully unconstrained 2-halo amplitudes. Maximum-posterior results from all 8 total tests are reported in Table \ref{tab:marginalized-asz}.

\begin{table}[!htbp]
    \centering
    {
    \tabulinesep=1.1mm
    \begin{tabu}{|c|c|c|}
    \hline
    $a_\obs{}$        & 0.1--12 Mpc                         & 0.1--5 Mpc                         \\ \hline
    Baryon           & $\fullRadiiBaryResult{}$            & $\cutRadiiBaryResult{}$            \\
    Baryon (1h only) & $\fullRadiiBaryOneHaloOnlyResult{}$ & $\cutRadiiBaryOneHaloOnlyResult{}$ \\
    NFW              & $\fullRadiiNfwResult{}$             & $\cutRadiiNfwResult{}$             \\
    NFW (1h only)    & $\fullRadiiNfwOneHaloOnlyResult{}$  & $\cutRadiiNfwOneHaloOnlyResult{}$  \\ \hline
    \end{tabu}
    }
    \caption{\label{tab:marginalized-asz} Marginalized maximum posteriors of the bias parameter $a_\obs{}$ for different models. Superscript and subscript values indicate distance to the $84$th and $16$th percentiles respectively. The baryon model is in all cases closer to the true value of $0$ than the equivalent NFW-only result.}
\end{table}

\section{Conclusions}

Our results demonstrate that modeling cluster weak-lensing signals only with an NFW profile results in a $\abstractNfwBias{}$ systematic overestimation of their masses via our stacking technique, whereas including a GNFW term to represent the deviation in density of baryons yields only $\abstractBaryBias{}$ overestimation. This demonstrates that, at the precision of the errors we use, our model is unbiased.

We believe the best way toward further validating our baryon model is through tests on additional simulations that use a variety of subgrid feedback mechanisms. Only in this manner can we properly establish the robustness of our model as an unbiased estimator of mass, independent of variance between simulations due to differing astrophysics. Another future direction would be to incorporate multi-wavelength cluster measurements of the baryon profile from X-ray, SZ, or fast radio burst observations. Such multi-wavelength information would provide empirical constraints on the baryon profile and should reduce the posteriors of our model's cluster mass calibrations, which are currently larger than the NFW-only model since we are marginalizing over these baryon profile parameters. The use of X-ray prior information has already been incorporated in the model proposed by Debackere, et al \citep{Debackere2021}.

Another aspect of our results that gives cause for comment is that our fit of the two-halo term peaks at no contribution despite the simulated data containing a visually obvious 2-halo regime. We briefly remarked on this in the previous section, but in more detail:
\begin{enumerate}
\item The errors on the large-radii bins are both fairly large, and highly correlated;
\item There are only 3 or 4 radial bins where the 2-halo term has much impact on the stack;
\item In the case of the NFW-only model, the best-fit region is overestimating the data at those radii, meaning the two-halo term will only make the model worse.
\
\end{enumerate}
In order to better constrain the two-halo term, larger radius bins are needed, or else smaller errors on the existing bins. Additionally, a summative approach ($1h + 2h$) has been shown to overpredict the transition regime \citep{oguri-hamana-2011}, which likely contributes to the suppression of the 2-halo term in our results. Adding a smooth truncation of the 1-halo term in the transition regime should also be explored.

In this work we have focused exclusively on the inclusion of baryons, but there are three additional corrections to our model needed for use on actual tSZ clusters: scatter and mass-dependence in the $M_\obs{}$--$M$ relation, and the mislocation of cluster centers when calculating the weak-lensing shear (miscentering). All contribute small but important effects to the ESD profiles which need to be taken into account to obtain an unbiased mass estimator. In our tests, we can ignore these effects since the simulation profiles were constructed with no miscentering, and the $M_\obs{}$ we input to our model were in fact the true masses. We did not attempt to construct analogous tests with miscentering and SZ-scatter or mass-dependence bias at this time, because both corrections render our model more computationally difficult to use. Including these corrections and testing them with simulations is a critical next step which is left for future work.

\section{Open Source Python Implementation}

We note that we have implemented the model described in this work as an open-source Python package, called \code{maszcal}. The source code for this package is available at \url{https://github.com/dylancromer/maszcal}. We will continue to use this package for our future work, where we expect to include further systematic corrections, perform additional tests, and continue to improve the software's documentation for others' use.

\section{Acknowledgements}

We thank Georgios Valogiannis, Ross Jennings, and Christopher Rooney for valuable discussions on various physical, mathematical, and computational aspects of this work. We thank Mathew S. Madhavacheril for valuable comments and suggestions.

\bibliographystyle{JHEP}
\bibliography{citations}

\appendix

\section{Testing the Accuracy of the Emulator} \label{app:emulator-accuracy}

In addition to testing emulator accuracy every time one is created for use in mass inference, we have performed an analysis of how the emulation errors change with respect to changing sample size and number of principal components. For this test, we use an arbitrary set of cluster masses and redshifts. We choose 400 halos whose masses are log-uniformly-distributed between $M_{500c} = 3 \times 10^{13} \, M_\odot$ and $3 \times 10^{14} \, M_\odot$. Redshifts are uniformly randomly distributed between $z = 0.2$ and $z=1$. Cluster weights are left to unity. This is not realistic, but it in fact results in larger emulator errors than performing the same tests on the simulated cluster mass and redshift distribution. Thus, this ad-hoc cluster sample is suitable for providing a conservative benchmark of the emulator performance.

In order to test the errors, we construct the emulator with the specified number of precalculated samples. We then choose randomly $1000$ additional parameter space samples, independent from the original sample, on which to test the emulator accuracy. We calculate the error via
\begin{equation}
    \frac{\delta D}{D_\mathrm{true}} = \frac{D_\mathrm{emulator} - D_\mathrm{true}}{D_\mathrm{true}}.
\end{equation}

\begin{figure}[!htbp]
    \centering
    \includegraphics[width=0.49\textwidth, keepaspectratio]{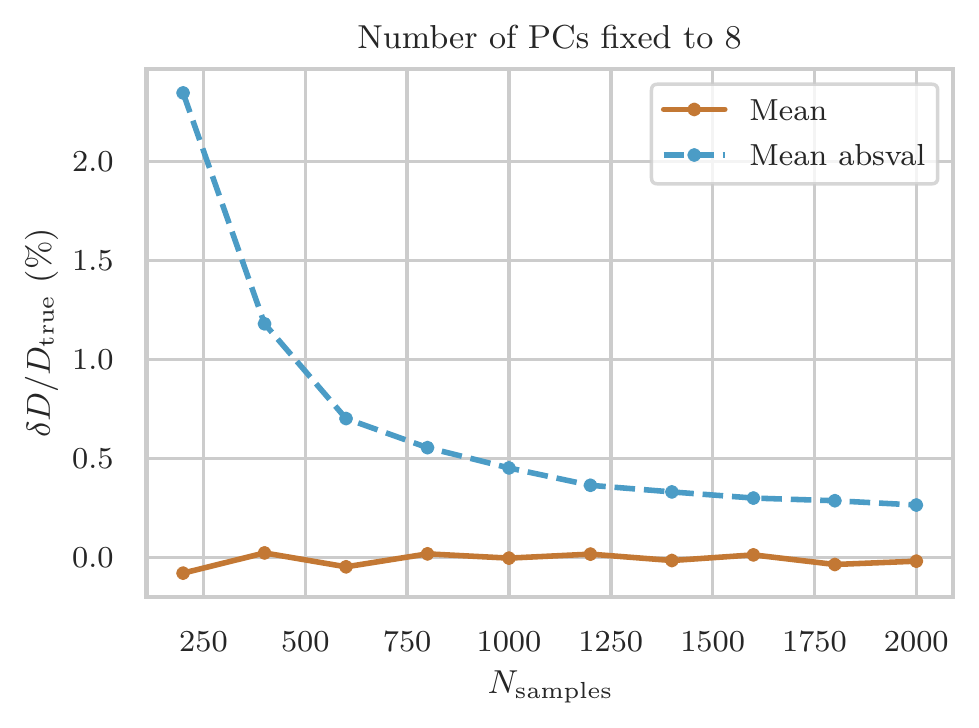}
    \includegraphics[width=0.49\textwidth, keepaspectratio]{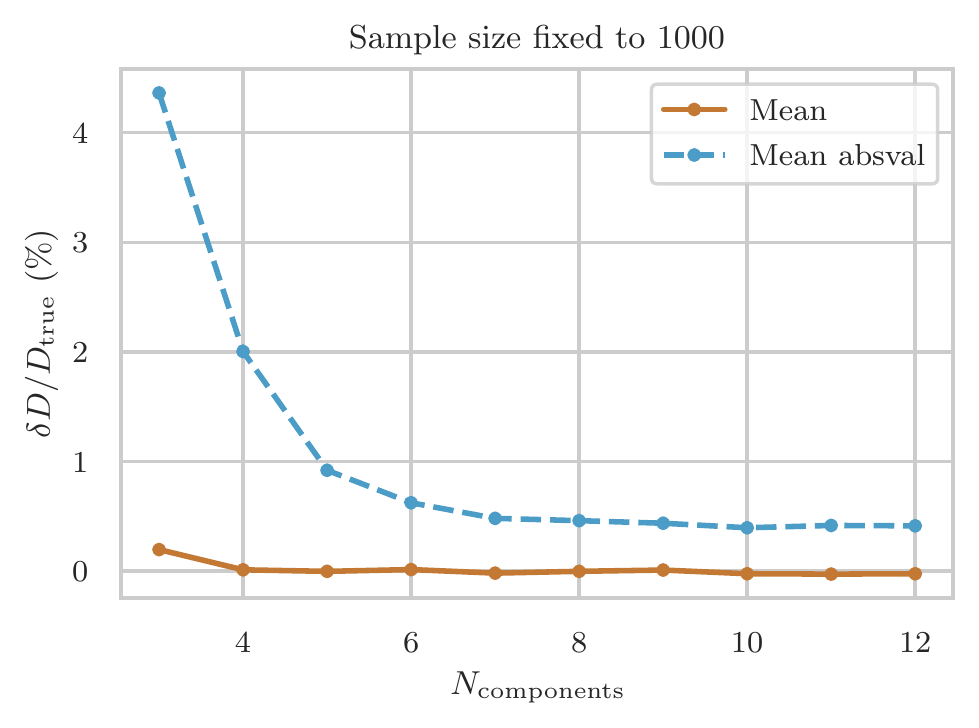}
    \caption{\label{fig:errors-by-samps-and-comps} Left: Mean error level of emulator as a function of sample size, using Latin hypercube samples and an emulator using 8 principal components. The error level is calculated by $(D_\mathrm{emulator} - D_\mathrm{true})/D_\mathrm{true}$. "Mean absval" indicates the absolute value of the error is taken before calculating the mean. Right: Ditto, but as a function of number of principal components instead of sample size. Here, sample size is fixed to 1000. This result indicates that the emulator is unbiased (mean error under $0.1\%$), and accurate (mean absolute error under $1\%$) when using 600 or more samples and 8 or more principal components.}
\end{figure}

In Figure \ref{fig:errors-by-samps-and-comps} is the mean error and mean absolute value error as a function of sample size and number of components respectively. The mean error is consistent with 0 in both cases. The mean absolute value error as a function of sample size shrinks to under $1\%$ by a sample size of $600$. Examining as a function of components reveals that only $5$--$6$ components are required before gains in accuracy diminish.

\begin{figure}[!htbp]
    \centering
    \includegraphics[width=0.8\textwidth, keepaspectratio]{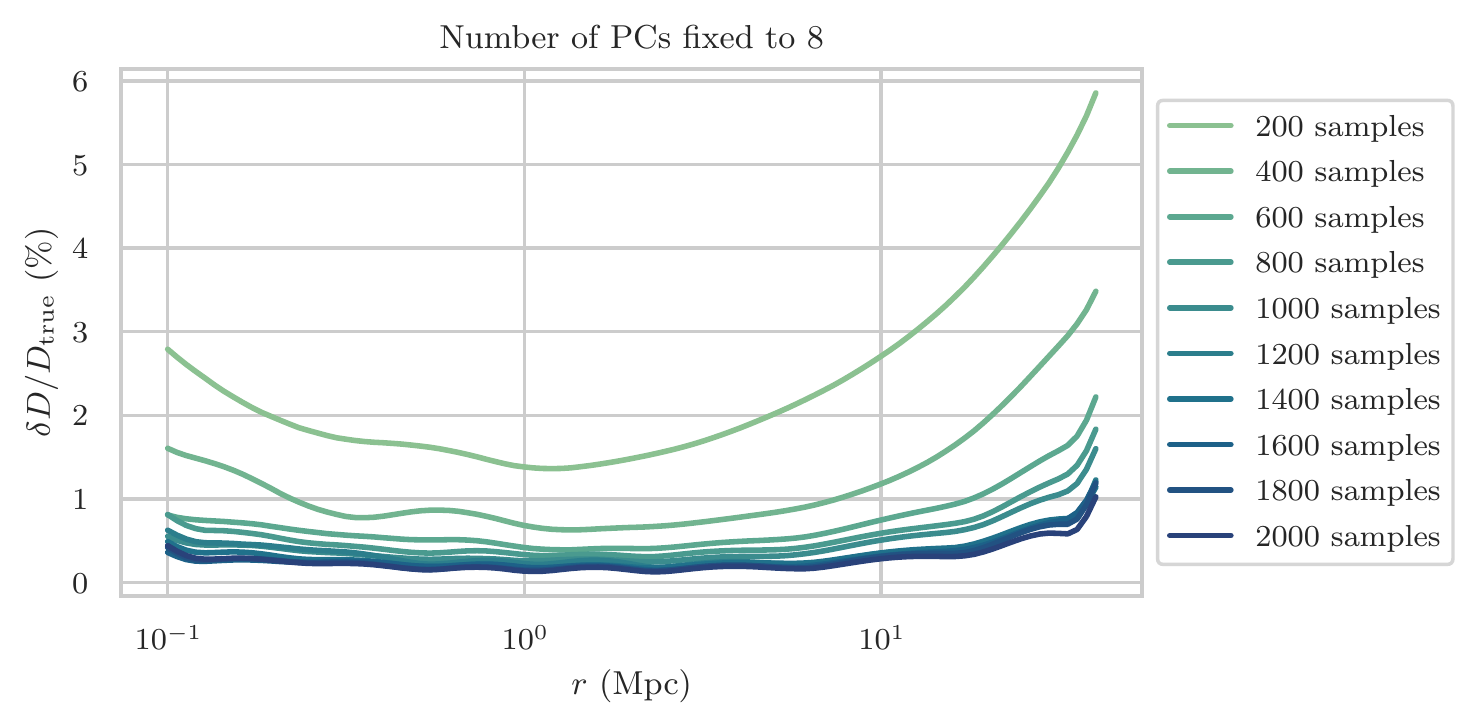}
    \caption{\label{fig:radial-avg-errors} Average emulator error as a function of radius for different Latin hypercube sample sizes. The errors are highest at the edges of the radius range, with the peak error creeping just above $1\%$ even when using 2000 samples. However, this peak in error level is at $\approx 20$ Mpc, above the maximum radius of the simulated data presented in this work.}
\end{figure}

We also examine the error averages as a function of sample size and radius in Figure \ref{fig:radial-avg-errors}. This indicates that the highest errors are at large radii.

\begin{figure}[!htbp]
    \centering
    \includegraphics[width=0.8\textwidth, keepaspectratio]{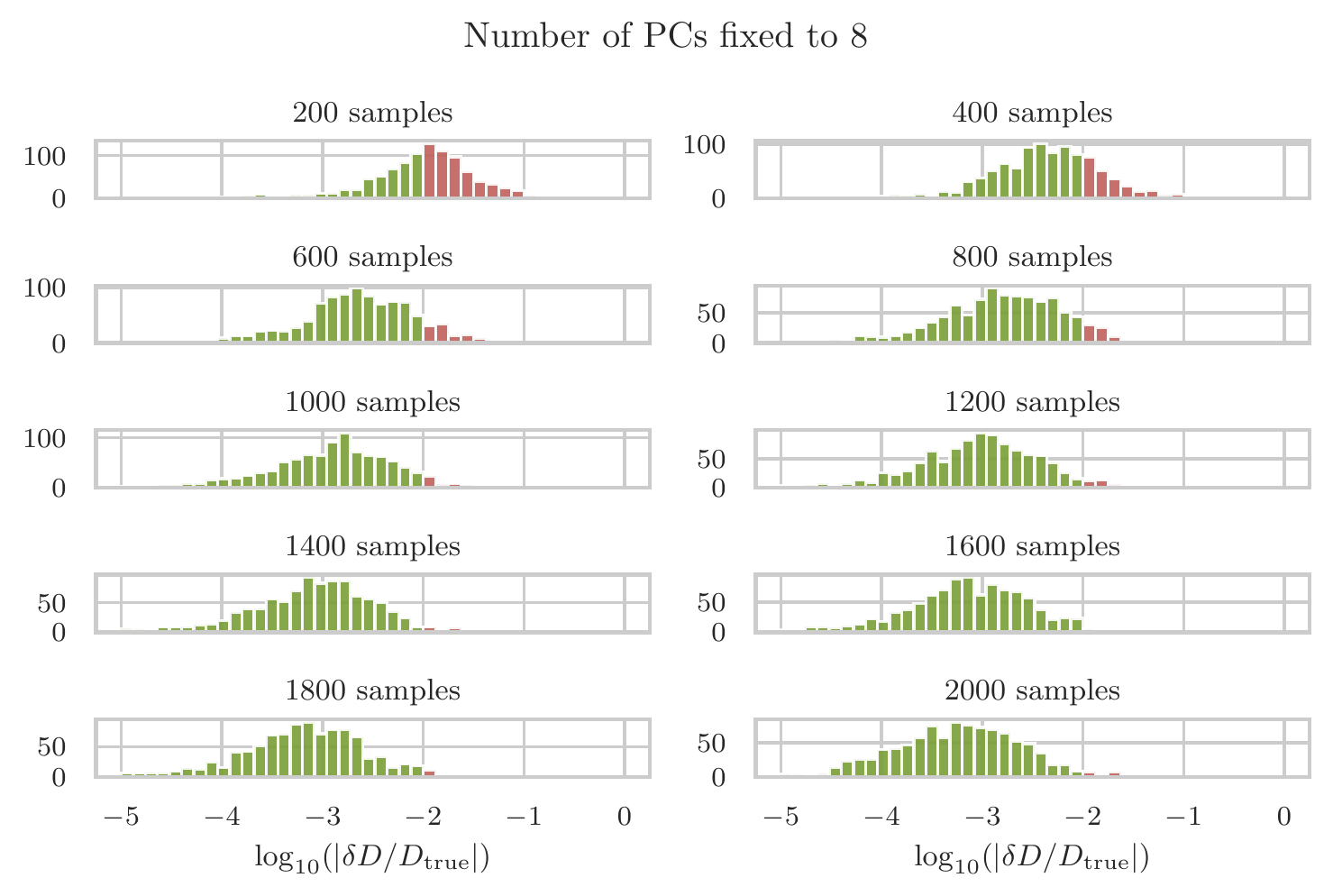}
    \caption{\label{fig:error-hists-by-sample-size} Histograms for base-10 logarithm of the absolute value emulation errors by sample size. In green are the histogram bins with errors below $1\%$; in red are the bins with errors above $1\%$. After using 1000 or more samples, the number of errors above $1\%$ becomes extremely low; we use 2000 samples for the main result of this work.}
\end{figure}

\begin{figure}[!htbp]
    \centering
    \includegraphics[width=0.8\textwidth, keepaspectratio]{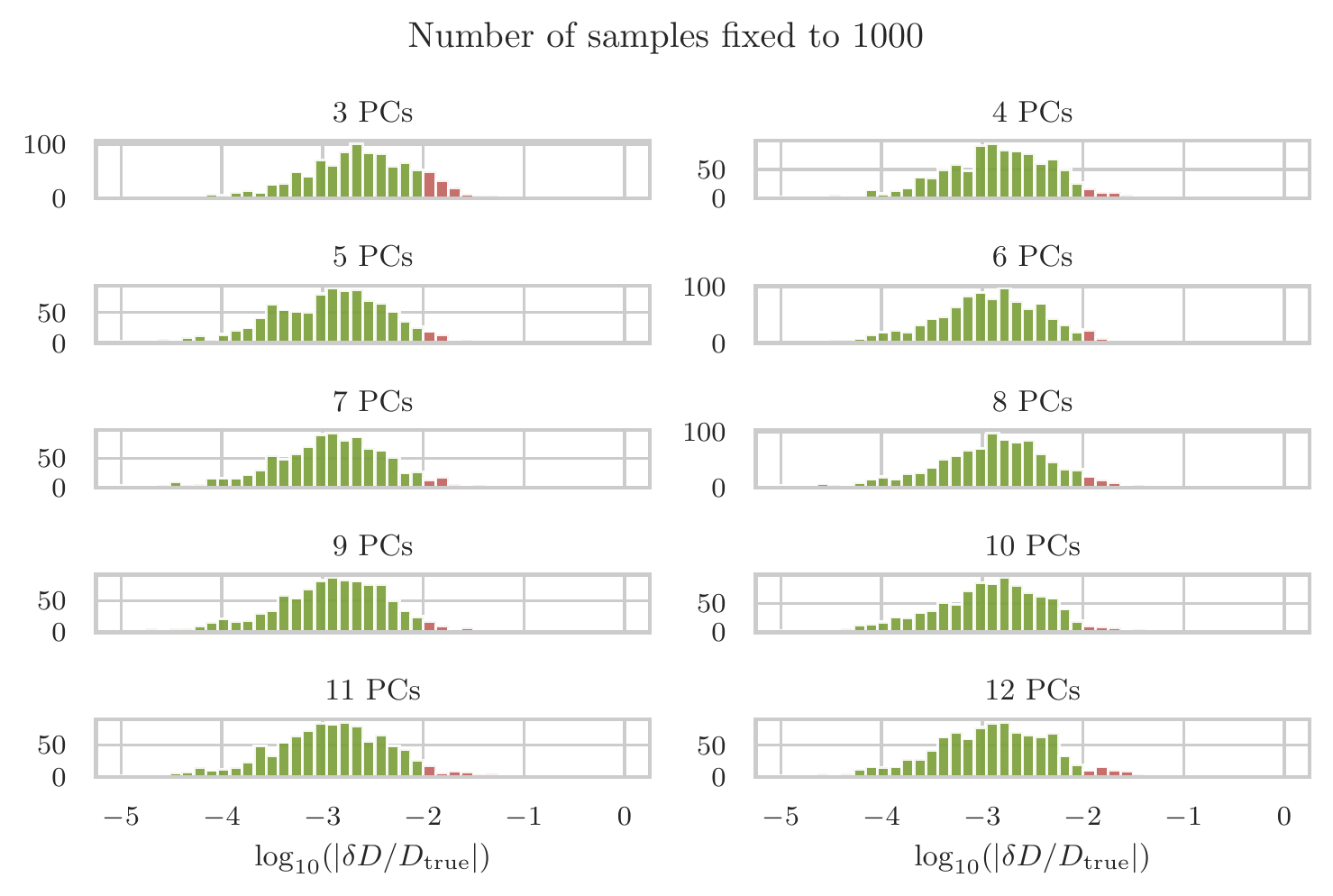}
    \caption{\label{fig:error-hists-by-components} Same as Figure \ref{fig:error-hists-by-sample-size}, but showing histograms for different numbers of principal components included. Once at least 6 are used, the impact of the number of principal components is negligible. In the main results, we use 10 or 8 components.}
\end{figure}

Lastly, we examine the histograms of the errors (specifically, the base-10 logarithm of the absolute value of the error) for each sample size and number of components, shown in Figures \ref{fig:error-hists-by-sample-size} and \ref{fig:error-hists-by-components}. These show explicitly the distribution of error magnitudes. When enough samples and components are used, very few errors are above $1\%$.

In summary, we find that if the emulator is provided sufficiently many samples, and uses at least 6 principal components or more, the error levels are very low and will not impact our results compared to using the bare model. For the primary result of this work, we use 2000 samples and 10 principal components for the baryon model, and 1200 samples and 8 principal components for the NFW-only model. Again, we reiterate that we also sample the parameter space randomly 1000 times during the creation of the main results, to ensure that the particular emulator used also has acceptably low errors. For each result in the main text, we find the emulator error levels to be lower than those shown in this appendix for $N_\mathrm{samples} = 2000$.

\section{Calculating Excess Surface Density Directly from a Density Profile} \label{app:rho_to_delta_sigma_calc}

We estimate the lensing shear statistic $\gamma$ using the excess surface density (ESD) $\Delta \Sigma(R)$. This is defined as
\begin{equation}
    \Delta \Sigma (R) = \ev{\Sigma(< R)} - \Sigma(R),
\end{equation}
where $\Sigma(R)$ is the projected surface density of the cluster, and $\SigmaAvg{R}$ is the average surface density within the radius $R$, given by
\begin{equation}
    \SigmaAvg{R} = \frac{\int_0^R \dd{r} r \Sigma(r)}{\int_0^R \dd{r} r} = \frac{2}{R^2} \int_0^R \dd{r} r \Sigma(r).
\end{equation}

The projected surface density itself is given by
\begin{align}
    \Sigma(R) &= \int_{-\infty}^{\infty} \dd{\ell} \rho(\losDist{})
              \\
              &= 2 \int_0^{\infty} \dd{\ell} \rho(\losDist{}), \label{eq:sd_form1}
\end{align}
where $\ell$ is the line-of-sight coordinate and $\losDist{} = \sqrt{\ell^2 + R^2}$ is the distance to the center of the profile along the integration axis. Expressing the integral in terms of $\losDist{}$ instead of $\ell$ gives
\begin{equation}
    \Sigma(R) = 2 \int_R^\infty \dd{\losDist{}} \frac{\losDist{} \rho(\losDist{})}{\sqrt{\losDist{}^2 - r^2}}. \label{eq:sd_form2}
\end{equation}

Thus in terms of $\rho$, the ESD is given by
\begin{equation}
    \Delta\Sigma(R) = \frac{4}{R^2} \int_0^R \dd{r} r \int_r^{\infty} \dd{\losDist{}} \frac{\losDist{} \rho(\losDist{})}{\sqrt{\losDist{}^2 - R^2}} - 2\int_R^{\infty} \dd{\losDist{}} \frac{\losDist{} \rho(\losDist{})}{\sqrt{\losDist{}^2 - r^2}}.
\end{equation}

Because $\rho$ depends on many parameters, we must perform this calculation many times for mass inference. The double integral is particularly costly from a computational standpoint, when dealing with large numbers of parameter space samples. Thus we seek to manipulate this form to one which is computationally easier to perform as a numerical integral.

We begin by focusing only on the double integral:
\begin{align}
\int_0^R \dd{r} r \int_r^{\infty} \dd{\losDist{}} \frac{\losDist{} \rho(\losDist{})}{\sqrt{\losDist{}^2 - r^2}} &= \int_0^R \dd{r} \int_r^\infty \dd{\losDist{}} \frac{r \losDist{} \rho(\losDist{})}{\sqrt{\losDist{}^2 - r^2}}.
\end{align}

As long as $\rho$ is continuous, we can switch the order of integration by noting that the region being integrated over in $r,\losDist{}$ coordinates is bounded by $r=\losDist{}$ for $\losDist{}<R$, and $r=R$ for $\losDist{} > R$. So if $\losDist{}$ is integrated first, the limits change to $\losDist{} = 0, \infty$, and $r=0$ and
\begin{equation}
    r = \begin{cases}
            R & \losDist{} > R
            \\
            \losDist{} & \losDist{} \leq R
        \end{cases}.
\end{equation}

Thus,
\begin{align}
    \int_0^R \dd{r} \int_r^\infty \dd{\losDist{}} \frac{r \losDist{} \rho(\losDist{})}{\sqrt{\losDist{}^2 - r^2}} 
        &= \int_0^R \dd{\losDist{}} \int_0^\losDist{} \dd{r} \frac{r \losDist{} \rho(\losDist{})}{\sqrt{\losDist{}^2 - r^2}}
        + \int_R^\infty \dd{\losDist{}} \int_0^R \dd{r} \frac{r \losDist{} \rho(\losDist{})}{\sqrt{\losDist{}^2 - r^2}}
        \\
        &= \int_0^R \dd{\losDist{}} 2 \losDist{}^2 \rho(\losDist{}) + \int_R^\infty \dd{\losDist{}} 2\losDist{} \qty(\losDist{} - \sqrt{\losDist{}^2 - R^2})\rho(\losDist{}),
\end{align}
where we can perform the integral over $r$ without specifying $\rho$. From here, we can calculate $\Delta \Sigma(R)$:
\begin{align}
    \Delta\Sigma(R) &= \frac{2}{R^2}\qty[\int_0^R \dd{\losDist{}} 2 \losDist{}^2 \rho(\losDist{}) + \int_R^\infty \dd{\losDist{}} 2\losDist{} \qty(\losDist{} - \sqrt{\losDist{}^2 - R^2})\rho(\losDist{})]
                        - 2\int_R^{\infty} \dd{\losDist{}} \frac{\losDist{} \rho(\losDist{})}{\sqrt{\losDist{}^2 - R^2}}
                    \\
                    &= \frac{2}{R^2} \qty[\int_0^R \dd{\losDist{}} 2 \losDist{}^2 \rho(\losDist{}) 
                        + \int_R^\infty \dd{\losDist{}} \qty(2\losDist{} \qty(\losDist{} - \sqrt{\losDist{}^2 - R^2}) - \frac{\losDist{} R^2}{\sqrt{\losDist{}^2 - R^2}})\rho(\losDist{})]
                    \\
                    &= \frac{2}{R^2} \qty[2 \int_0^R \dd{\losDist{}} \losDist{}^2 \rho(\losDist{})
                        + \int_R^\infty \dd{\losDist{}} \losDist{} \qty(2\losDist{} + \frac{R^2 - 2\losDist{}^2}{\sqrt{\losDist{}^2 - R^2}})\rho(\losDist{})].
\end{align}

To compress the second term to an integral over a finite coordinate region, we can change coordinates to $\theta$, the complement of the angle from the line of sight-axis, by $x = R/\cos{\theta}$. In terms of these coordinates, the second term (sans the prefactor $2/R^2$) becomes
\begin{align}
    \qquad& R^3 \int_0^{\pi/2} \dd{\theta} \frac{\tan{\theta}}{\cos^2{\theta}}
    \qty(\frac{2}{\cos{\theta}} + \frac{1 - 2/\cos^2{\theta}}{\sqrt{1/\cos^2{\theta} - 1}}) \rho(R\sec{\theta})
    \\
    &= R^3 \int_0^{\pi/2} \dd{\theta} \frac{\tan{\theta}}{\cos^3{\theta}} \qty(2 - \frac{1+\sin^2{\theta}}{\sin{\theta}})\rho(R\sec{\theta})
    \\
    &= R^3 \int_0^{\pi/2} \dd{\theta} \frac{2 \rho(R\sec{\theta})}{\cos(2\theta) - 4 \sin{\theta} - 3}.
\end{align}

Then, the final expression for $\Delta \Sigma(R)$ is
\begin{equation}
    \Delta \Sigma(R) = \frac{4}{R^2} \int_0^R \dd{\losDist{}} \losDist{}^2 \rho(\losDist{}) - 4R \int_0^{\pi/2} \dd{\theta} \frac{\rho(R\sec{\theta})}{4\sin{\theta} + 3 - \cos(2\theta)}
\end{equation}

\section{Scatter in the Observable-Mass Relation} \label{app:scatter}
The observable-mass relation for, e.g., the tSZ Compton-$Y$, contains scatter. This scatter is important to model as it can modify the stacked lensing signal by anywhere from $1\%$ to $10\%$. It is easy to model scatter in our stacking methodology. Suppose we use a lognormal distribution given by
\begin{equation}
    p(\mu_\obs|\mu) = \frac{1}{\sqrt{2\pi}\sigmaobs} \exp[-\frac{\qty(\mu_\obs{} - b_\obs{} \mu - a_\obs{})^2}{2\sigmaobs{}^2}] \label{eq:sz-mass-relation-scatter}
\end{equation}
to model the relation between $M$ and $\Mobs{}$. Here, $b_\obs{}$ would control a mass-dependent exponent in the bias, and $\sigmaobs{}$ the magnitude of the scatter. To capture this, the matching model can be averaged over the possible masses that could have given rise to the observed $\Mobs{}$'s:
\begin{align}
    \bar D(r) &= \sum_i w_i \int \dd{\mu} D\qty(r, z_i, e^\mu, \vb*{\theta}) p_i(\mu|\mu_{SZ{}\, i}; \, a_\obs{}, b_\obs{}, \sigmaobs{})
              \\
              &= \sum_i w_i \int \dd{\mu} D\qty(r, z_i, e^\mu, \vb*{\theta}) p_i(\mu) p_i(\mu_{{SZ}\, i}|\mu; \, a_\obs{}, b_\obs{}, \sigmaobs{}) / p_i(\mu_{\obs{}\, i}),
\end{align}
where 
\begin{equation}
    p_i(\mu_{\obs{}\, i}) = \int \dd{\mu} p_i(\mu) p_i(\mu_{{SZ}\, i}|\mu; \, a_\obs{}, b_\obs{}, \sigmaobs{}),
\end{equation}
where $p_i(\mu_{{SZ}\, i}|\mu; \, a_\obs{}, b_\obs{}, \sigmaobs{})$ is as given in Equation \ref{eq:sz-mass-relation-scatter} but evaluated at the $i$th SZ mass, and $p_i(\mu)$ is the halo mass function evaluated at the $i$th cluster redshift:
\begin{align}
    p_i(\mu) &= \eval{\qty[\pdv{V}{z}{\Omega} \dv{n}{\mu} \bigg/ \int \dd{\mu'} \pdv{V}{z}{\Omega} \dv{n}{\mu'}]}_{z=z_i} \\
    &= \eval{\qty[\dv{n}{\mu} \bigg/ \int \dd{\mu'} \dv{n}{\mu'}]}_{z=z_i}.
\end{align}

\end{document}